\documentclass[fleqn,useAMS,usenatbib]{mnras}
\usepackage[flushleft]{threeparttable}
\usepackage{soul}
\usepackage{ragged2e}
\usepackage[usenames]{color}

\definecolor{green2}{rgb}{0.15,0.6,0.05}
\definecolor{rust}{rgb}{0.7,0.1,0.1}

\usepackage{adjustbox}
\usepackage{orcidlink}
\usepackage[super]{nth}
\usepackage{hyperref}
\usepackage{longtable}
\usepackage{stfloats} 
\usepackage[caption=false]{subfig} 

\DeclareRobustCommand{\VAN}[3]{#2} 
\title[AT~2016blu]
{Recurring outbursts of the supernova impostor AT~2016blu in NGC~4559}

\author[M. Aghakhanloo et al.]{Mojgan Aghakhanloo \orcidlink{0000-0001-8341-3940},$^{1}$\thanks{E-mail:
aghakhanloo@arizona.edu} Nathan Smith \orcidlink{0000-0001-5510-2424},$^1$ 
Peter Milne,$^1$ Jennifer E. Andrews \orcidlink{0000-0003-0123-0062},$^{2}$ \newauthor
Schuyler D. Van Dyk \orcidlink{0000-0001-9038-9950},$^{3}$ 
Alexei V. Filippenko \orcidlink{0000-0003-3460-0103},$^{4}$ 
Jacob E. Jencson
\orcidlink{0000-0001-5754-4007},$^{5}$ 
Ryan M. Lau
\orcidlink{0000-0003-0778-0321},$^{6}$ \newauthor
David J. Sand \orcidlink{0000-0003-4102-380X},$^{1}$ 
Samuel Wyatt \orcidlink{0000-0003-2732-4956},$^{7}$ 
WeiKang Zheng \orcidlink{0000-0002-2636-6508}$^{4}$\\ 
$^1$ Steward Observatory, University of Arizona, 933 N. Cherry Ave., Tucson, AZ 85721, USA  \\ 
$^2$ Gemini Observatory, 670 N. Aohoku Place, Hilo, HI 96720, USA \\ 
$^3$ Caltech/IPAC, Mailcode 100-22, Pasadena, CA 91125, USA\\
$^4$ Department of Astronomy, University of California, Berkeley, CA 94720-3411, USA\\
$^5$ Department of Physics and Astronomy, The Johns Hopkins University, Baltimore, MD 21218, USA\\
$^6$ NSF’s NOIRLab, 950 N. Cherry Avenue, Tucson, AZ 85719, USA\\
$^7$ Department of Physics, University of Washington, Seattle, WA 98195, USA}

\begin{document}
\pagerange{\pageref{firstpage}--\pageref{lastpage}} \pubyear{2023}
\maketitle
\label{firstpage}

\begin{abstract}
We present the first photometric analysis of the supernova (SN) impostor AT~2016blu in NGC~4559. This transient was discovered by the Lick Observatory Supernova Search in 2012 and has continued its outbursts since then. Optical and infrared photometry of AT~2016blu reveals at least 19 outbursts in 2012--2022. Similar photometry from 1999--2009 shows no outbursts, indicating that the star was relatively stable in the decade before discovery. Archival {\it Hubble Space Telescope} observations suggest that the progenitor had a minimum initial mass of $M \ga 33$~M$_{\odot}$ and a luminosity of $L \ga 10^{5.7}$~L$_{\odot}$. AT~2016blu's outbursts show  irregular variability with multiple closely spaced peaks having typical amplitudes of 1--2 mag and durations of 1--4 weeks. While individual outbursts have irregular light curves, concentrations of these peaks recur with a period of $\sim 113 \pm 2$~d. Based on this period, we predict times for upcoming outbursts in 2023 and 2024. AT~2016blu shares similarities with SN~2000ch in NGC~3432, where  outbursts may arise from periastron encounters in an eccentric binary containing a luminous blue variable (LBV). We propose that AT~2016blu's outbursts are also driven by interactions that intensify around periastron in an eccentric system.  Intrinsic variability of the LBV-like primary star may cause different intensity and duration of binary interaction at each periastron passage.  AT~2016blu also resembles the periastron encounters of $\eta$~Carinae prior to its Great Eruption and the erratic pre-SN eruptions of SN~2009ip. This similarity and the onset of eruptions in the past decade hint that AT~2016blu may also be headed for a catastrophe, making it a target of great interest.

\end{abstract}

\begin{keywords}
stars: individual: AT~2016blu -- stars: massive -- stars: variables: general -- galaxies: NGC~4559.

\end{keywords}

\section{INTRODUCTION}\label{sec:intro}

``Supernova (SN) impostors" are a class of eruptive transient sources found alongside supernovae (SNe) in modern surveys, although they are generally less luminous than SNe and usually have narrow ($\leq 1000$~km~s$^{-1}$) H emission lines \citep{V00,S11}.  SN impostors display a wide variety of peak luminosity and light-curve shape, and are usually associated with non-terminal eruptions of evolved massive stars like luminous blue variables (LBVs), based on comparison with giant eruptions of LBVs like $\eta$~Carinae and P~Cygni where the stars are known to have survived. However, the progenitors and physical causes of SN impostor eruptions may be diverse. In the case of AT~2016blu, the transient source discussed in this paper, the star survived its outbursts (and is therefore a likely SN impostor) because it continued to have multiple additional outbursts up to the present time.

\begin{figure*}
  \centering
  \subfloat[\label{fig:ds9image}]{\raisebox{-0.5\height}{\includegraphics[width=0.54\textwidth]{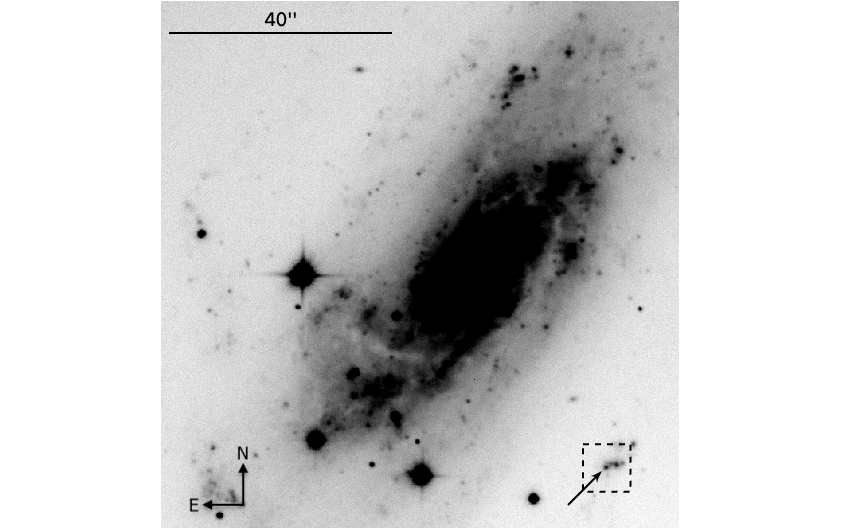}}}
  \subfloat[\label{fig:precursor}]{\raisebox{-0.5\height}{\includegraphics[width=0.355\textwidth]{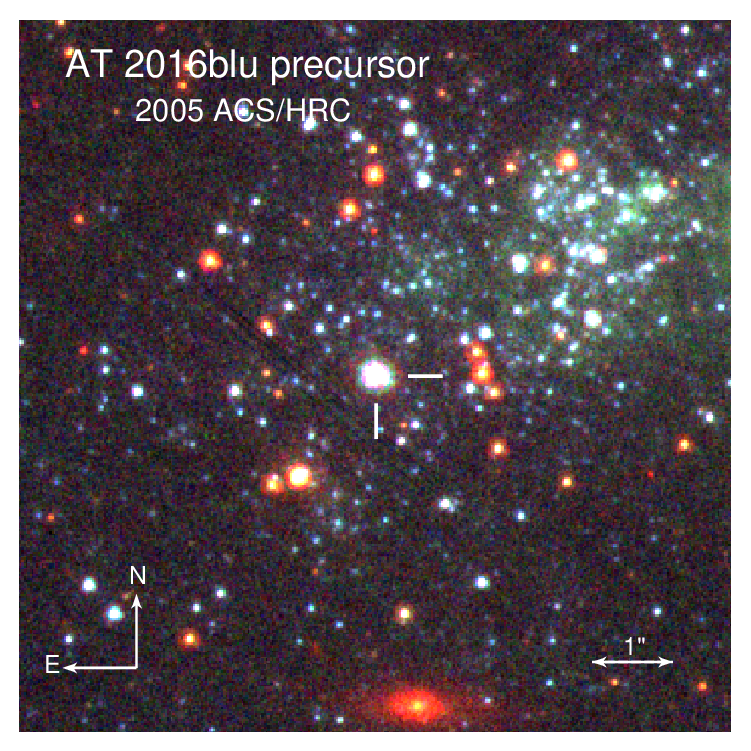}}}
  \caption{The left panel shows the Kuiper $R$-band image of AT~2016blu obtained on 2018 March 20. The position of AT~2016blu is marked with an arrow. The area inside the dashed square is shown in more detail in the right panel. The right panel presents a colour composite of a portion of the 2005 {\it HST} ACS/HRC image mosaic (F435W, blue; F555W, green; F814W, red) showing the AT~2016blu progenitor, indicated by the tick marks. The object was originally identified by \citet{V12}.}
  \label{fig:images2}
\end{figure*}

AT~2016blu (also known as NGC~4559OT) was discovered as a new transient source on 2012 January 12 (UTC dates are used throughout this paper) by \citet{K12} during the course of the Lick Observatory Supernova Search \citep[LOSS;][]{L02,F01} with the 0.76~m Katzman Automatic Imaging Telescope (KAIT).
AT~2016blu is located about 71{\arcsec} west and 103{\arcsec} south of the centre of the galaxy NGC~4559 (see Fig.~\ref{fig:ds9image}), in an H~{\sc ii} region known as NGC~4559C (or IC 3550). NGC~4559C is at a distance of 8.91~Mpc ($\pm 0.19$ statistical, $\pm 0.29$ systematic) \citep{Mc17}, with Galactic reddening of $E(B-V)=0.015$~mag \citep{Schlafly2011}.  However, \citet{Sh14} suggested that AT~2016blu may not be associated with NGC~4559 itself. Given the spatial location and measured recession velocity, they speculated that AT~2016blu is associated with a nearby faint satellite galaxy SDSS J123551.86+275556.9. \cite{K12} reported that an early-time optical spectrum of AT~2016blu resembled that of the SN impostor SN~2009ip, leading them to suggest  that AT~2016blu's 2012 outburst is most likely the eruption of an LBV. AT~2016blu's spectrum exhibits hydrogen Balmer and He~{\sc i} lines with complex P~Cygni profiles. The emission lines have both narrow and broad components. The narrow lines may arise from slow-moving circumstellar material (CSM), but they might also be contaminated by the emission of an underlying H~{\sc ii} region.  Unresolved [S~{\sc ii}] and [O~{\sc iii}] emission lines in the spectrum were attributed to the H~{\sc ii} region surrounding AT~2016blu \citep{Sh14}. (Note, on the other hand, that in a spectrum we present later in this paper, the narrow H$\alpha$ emission persists even though there is little or no contamination from an H~{\sc ii} region, judging by the absence of [S~{\sc ii}] emission in the same spectrum.) Similarly, ground-based photometric data of AT~2016blu might also be contaminated by continuum emission from the surrounding stellar association. 

As we discuss later in this paper, the recurring outbursts of AT~2016blu appear somewhat similar to the series of quasiperiodic outbursts observed in the SN impostor SN~2000ch \citep{W04,P10}. We recently discussed in detail the long-term light curve of SN~2000ch \citep[hereafter Paper I;][]{AA22}. SN~2000ch experienced at least 23 known outbursts in 2000--2022. Quasiperiodicities observed in the light curve along with diversity in the properties of outbursts suggest that SN~2000ch's eruptions are caused by interaction around times of periastron in an eccentric binary system. This scenario involves the interplay of two different variability phenomena. One is the intrinsic S~Dor-like variability of the LBV (with a timescale of years), and the other is the observed bright outbursts/eruptions of SN~2000ch (typically lasting a few to several weeks), which are presumably driven by the interaction between the two stars at periastron in the eccentric binary system. (In this work, the terms ``eruption" and ``outburst" are used interchangeably.) Given recent evidence that most LBVs are the products of binary evolution \citep{ST15,A17}, it may be unsurprising if the apparent variability of AT~2016blu also arises in an eccentric, interacting binary system, similar to that of SN~2000ch. 

The eruptions of AT~2016blu and SN~2000ch have photometric properties similar to the rapid brightening and fading observed in pre-SN eruptions of the well-known SN impostor SN~2009ip \citep{S10}. SN~2009ip provides the most direct link between LBVs and Type IIn SNe \citep{M13}. Recently, \citet{S22} confirmed a terminal core-collapse SN scenario for the dramatic 2012 event in SN 2009ip, because
the luminous progenitor star is now gone. Recent evidence also confirms more terminal SN explosions with LBV-like progenitors in other SNe~IIn \citep{B22, J22}. Given the similarities to SN~2009ip, one might naturally wonder if AT~2016blu and SN~2000ch will soon undergo SN explosions, making them targets of great interest.

In this paper, we present the first photometric investigations and period analyses of AT~2016blu. Our main goal is to document major eruptions of AT~2016blu and search for any degree of periodicity similar to that of SN~2000ch. 
First, in Section~\ref{sec:obs} we describe the observations. In Section~\ref{sec:LC}, we analyse the major outbursts of AT~2016blu. We discuss AT~2016blu's infrared variability and study its colour evolution in Section~\ref{sec:Wavelength}. Section~\ref{sec:Priodicity} investigates the possible periodic variability of AT~2016blu. In Section~\ref{sec:Precursor}, we analyse {\sl Hubble Space Telescope\/} ({\sl HST}) data to constrain the mass and luminosity of AT~2016blu's progenitor. Section~\ref{sec:binary} speculates about the underlying mechanism causing these outbursts, and similarities with SN~2000ch and SN~2009ip. We summarise our conclusions in  Section~\ref{sec:conclusion}.

\section{OBSERVATIONS}\label{sec:obs}
\subsection{Ground-Based Photometry}

We obtained optical photometry of AT~2016blu using the Katzman Automatic Imaging Telescope \citep[KAIT;][]{F01} at Lick Observatory from January 1999 to the present epoch. The data-reduction process is described in Paper I. The unfiltered KAIT photometry of AT~2016blu is summarised in Table~\ref{tab:KAIT}. Unfiltered KAIT magnitudes are similar to the standard $R$ passband \citep{L03}. Upper limits derived from  KAIT images are available in the online material.

\begin{figure}
\includegraphics[width=0.45\textwidth]{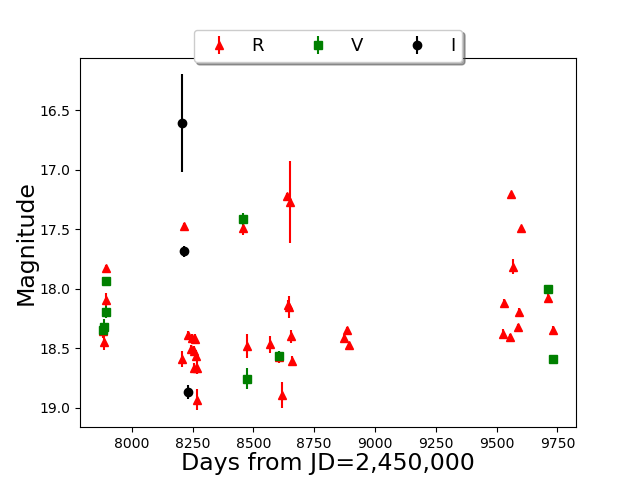}
\caption{The light curve of AT~2016blu using Super-LOTIS photometry. See Table~\ref{tab:Super-LOTIS}.}\label{fig:Super-LOTIS}
\end{figure}

Optical photometry of AT~2016blu was also obtained with
the 0.6~m robotic Super-LOTIS telescope \citep[Livermore Optical Transient Imaging System;][]{W08} on Kitt Peak,
from 2017 May 6 up to the present epoch. Super-LOTIS optical magnitudes and associated uncertainties are given in Table~\ref{tab:Super-LOTIS}. Fig.~\ref{fig:Super-LOTIS} shows the multiband light curve of AT~2016blu observed by Super-LOTIS. In addition, AT~2016blu was observed using the MONT4K CCD imager on the 61-inch Kuiper telescope on Mt. Bigelow, AZ; the Kuiper photometry is presented in Table~\ref{tab:Kuiper}.

\begin{figure}
\includegraphics[width=0.45\textwidth]{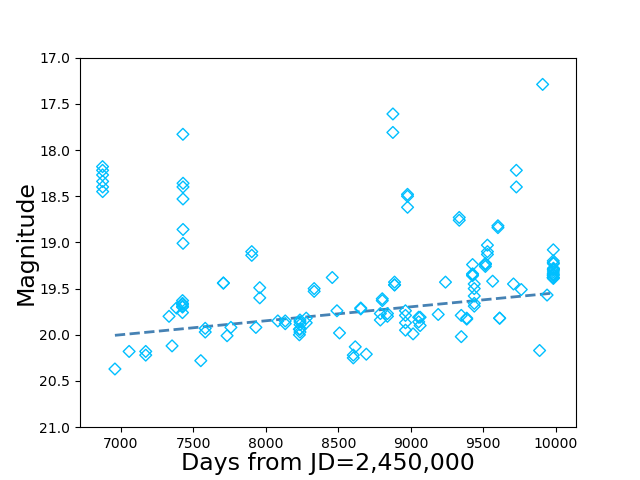}
\caption{{\it Gaia} light curve of AT~2016blu. The blue dashed line represents the linear regression fit to the data with magnitudes fainter than 19.3. The fit indicates that the baseline magnitude of the source is increasing by $\sim 0.5$~mag over a span of $\sim 8$~yr. }\label{fig:Gaia}
\end{figure}

AT~2016blu was detected by the {\it Gaia} space telescope \citep{G16}. {\it Gaia} Photometric Science Alerts \citep{H21} report sources that show significant brightness changes. Gaia16ada is the first SN impostor eruption reported by {\it Gaia}, and its position coincides with that of AT~2016blu. Fig.~\ref{fig:Gaia} shows the {\it Gaia} light curve of AT~2016blu. The blue dashed line represents the linear regression fit to the base magnitude ($>19.3$~mag), indicating an increase of $\sim 0.5$~mag in the base magnitude over a period of $\sim 8$~yr. {\it Gaia} $G$-band photometry of Gaia16ada is available in the {\it Gaia} Science Alerts system\footnote{\url{ http://gsaweb.ast.cam.ac.uk/alerts/home}}. 

We also retrieved available photometry of AT~2016blu from the Zwicky Transient Facility \citep[ZTF;][]{B19} public surveys.
ZTF data are available online\footnote{\url{https://irsa.ipac.caltech.edu/Missions/ztf.html}}. We collect the sources within 1{\arcsec} of the position of AT~2016blu from ZTF DR16. Data from the original Palomar Transient Factory \citep[PTF;][]{L09} are not used because the coordinates are several arcseconds off. We suspect that the source in the PTF photometry is dominated by light from the nearby stellar association, not the transient. 

AT~2016blu was observed by the ATLAS Project \citep[Asteroid Terrestrial-impact Last Alert System;][]{T18}.  ATLAS data are available in the ATLAS Forced Photometry server\footnote{\url{https://fallingstar-data.com/forcedphot/}}. Data with poor quality and large error bars are excluded from ZTF and ATLAS data; see Paper I for more details. We should note that because AT~2016blu is buried in an H~{\sc ii} region and a stellar association, we used ATLAS forced photometry on the difference images instead of forced photometry on the target images.

We include various discovery data from the ``Bright Supernovae" website\footnote{\url{https://www.rochesterastronomy.org/supernova.html}} maintained by David W. Bishop. 
This website compiles data from amateur astronomers whose observations are collected in various types of filters (or unfiltered). Later in Section~\ref{sec:LC}, we only use observations that are reported as discovery. The Discovery report specifies that AT~2016blu was bright at the time and there should be outburst activity around the reported time. However, since AT~2016blu exhibits multiple narrow peaks (see Section~\ref{sec:LC} for more details), photometry gathered on the Bright Supernovae website might correspond to the fluctuations before or after the main event. If these discoveries are not supplemented with more data, the precise time of the outburst remains uncertain.

Infrared (IR) data were obtained with the {\it Spitzer Space Telescope},
as described below.

\subsection{{\sl HST} Observations}
\label{sec:HSTobs}

The site of AT~2016blu was serendipitously observed on three occasions with {\sl HST\/} in the decade prior to the 2012 outburst: on 2001 May 25 (GO-9073, PI: J.~Bregman) with the 
Wide-Field Planetary Camera 2 (WFPC2) in bands F450W, F555W, and F814W, each with 2000~s exposure time; on 2005 March 8 (GO-10214, PI R.~Soria) with the 
Advanced Camera for Surveys (ACS)/High Resolution Channel (HRC) in F435W, F555W, F814W, F502N, and F658N, each with 2400~s exposure time; and on 2005 March 9 (GO-10214) with the ACS/Wide Field Channel (WFC) in F606W (5920~s) and F550M (5581~s). All of these publicly-available archival data were obtained from the Mikulski Archive for Space Telescopes (MAST) portal. We first processed the individual pipeline-processed exposure frames with Astrodrizzle \citep{STSCI2012}, before running them through the {\sl HST}-optimised photometry routine Dolphot \citep{Dolphin2016}; a major advantage of undertaking the former is the identification and masking of cosmic-ray hits in the frames, removing the effect of these artefacts on the Dolphot aperture-correction step.

Our analysis here was greatly simplified, since \citet{V12} had already located the progenitor in these {\sl HST\/} data, which we show in Fig.~\ref{fig:precursor}. The object is in the same complex region of the galaxy as the optical counterpart of the ultraluminous X-ray source X7 studied by \citet{Soria2005}, although $\sim 12${\arcsec} to the southeast (cf.~their Fig.~2). It is striking that the AT~2016blu progenitor is one of the brightest objects in the field in these pre-outburst images. Interestingly, \citet{Soria2005} estimate that the evolved stars in the surrounding region (mostly blue and red supergiants) have luminosities corresponding to initial masses of 10--15~M$_{\odot}$, implying ages of $\sim 20$~Myr for the population.
The field of view in Fig.~\ref{fig:precursor} 
includes an 8{\arcsec} $\times$ 8{\arcsec} region, showing the stellar population within a radius of 175~pc.  
{\sl HST\/} raw magnitudes are given in Table~\ref{tab:HST}. In Section~\ref{sec:Precursor}, we use the {\sl HST\/} data to constrain properties of the AT~2016blu progenitor.

\subsection{Infrared Photometry}

The site of AT~2016blu was also imaged by the {\it Spitzer Space Telescope} ({\it Spitzer}) at wavelengths of 3.6 and 4.5~$\mu$m ({\it Spitzer}/IRAC Bands 1 and 2). To measure the brightness of the object, we used a 2 native-pixel (radius) aperture of 2.4{\arcsec}. We used a background annulus ranging from 2 to 6 native pixels (2.4--7.2{\arcsec}) to measure and subtract the background emission. We applied aperture-correction factors of 1.2132 and 1.2322 for the Ch1 and Ch2 filters, respectively. These values are consistent with the correction factors recommended in the {\it Spitzer}/IRAC manual for the chosen aperture and background annulus radii. The {\it Spitzer} photometry can be found in Table~\ref{tab:spitzer}.

The angular resolution of the {\it Spitzer} imaging is significantly lower than the {\sl HST\/} observations, and the {\it Spitzer} photometry therefore necessarily includes contamination from some of the neighbouring stars seen in Fig.~\ref{fig:images2}b.  We are mostly interested in changes in the IR excess emission that may be associated with outbursts in the past decade.  Therefore, we also subtracted a baseline flux corresponding to the epoch in 2004 when the source was at its faintest recorded magnitude in the {\it Spitzer} bands before the first outburst in 2012, and the differences in magnitudes are discussed below.  Similarly, the last column in Table~\ref{tab:spitzer} includes the Ch1/Ch2 colour after subtracting this 2004 baseline flux level.

\subsection{Spectroscopy}
\label{sec:Spectrum}
\begin{figure}
\includegraphics[width=0.47\textwidth]{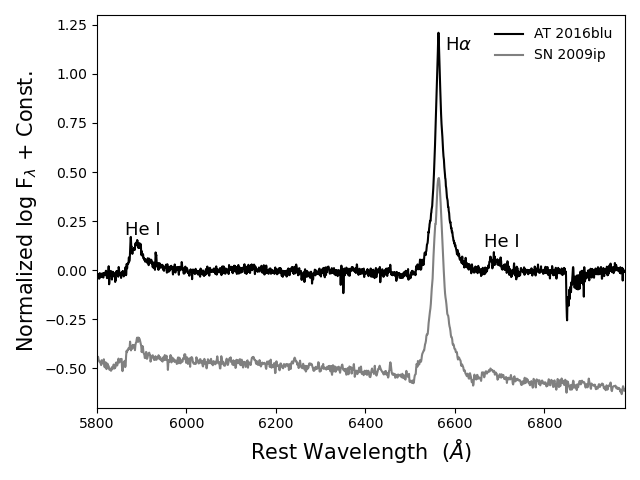}
\caption{MMT Blue Channel spectrum of AT~2016blu obtained on 2019 June 4, coincident in time with AT~2016blu's \nth{11} outburst; see Fig.~\ref{fig:RecentZoomed9}. H$\alpha$ does not show a P~Cygni profile and its Lorentzian FWHM is $\sim 415$~km~s$^{-1}$; correcting by quadrature for the instrumental resolution of $\sim 80$~km~s$^{-1}$ with the 1200 line~mm$^{-1}$ grating, the FWHM is $\sim 410$~km~s$^{-1}$.
The He~{\sc i} lines at 5876~\AA\ and 6680~\AA\ are also visible.  The day 25 spectrum of SN~2009ip from \citet{S10} is shown for comparison.}\label{fig:spectrum}

\end{figure}

An optical spectrum of AT~2016blu (Fig.~\ref{fig:spectrum}) was obtained on 2019 June 4 with the Blue Channel spectrograph on the 6.5~m MMT Observatory on Mount Hopkins, Arizona, coincident in time with AT~2016blu's \nth{11} outburst (see Section~\ref{sec:LC} for more details). It exhibits a very strong, relatively narrow emission line of H$\alpha$ with a Lorentzian full width at half-maximum intensity (FWHM) of $\sim 410$~km~s$^{-1}$ (corrected for the instrumental resolution). Weaker emission lines from He~{\sc i} at $\lambda$5876 and $\lambda$6680~\AA\ are also visible. In Fig.~\ref{fig:spectrum}, AT~2016blu’s spectrum is compared with that of the 2009 pre-SN outburst of SN~2009ip from \citet{S10}. The spectra are very similar, and the $\sim 550$~km~s$^{-1}$ FWHM of the H$\alpha$ line in SN~2009ip \citep{S10} is comparable to that of AT~2016blu reported here. A much more detailed analysis of AT~2016blu's spectral evolution will be discussed in a forthcoming paper.

\section{The Light Curve}\label{sec:LC}
\begin{figure*}
\subfloat[]{
  \includegraphics[width=0.8\textwidth]{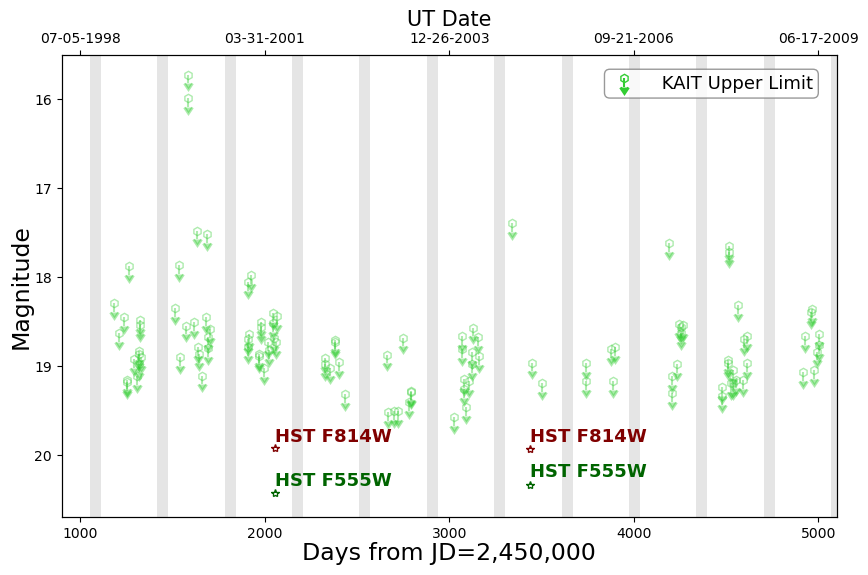}
  \label{fig:LCUL}}\\
\subfloat[]{
  \includegraphics[width=0.8\textwidth]{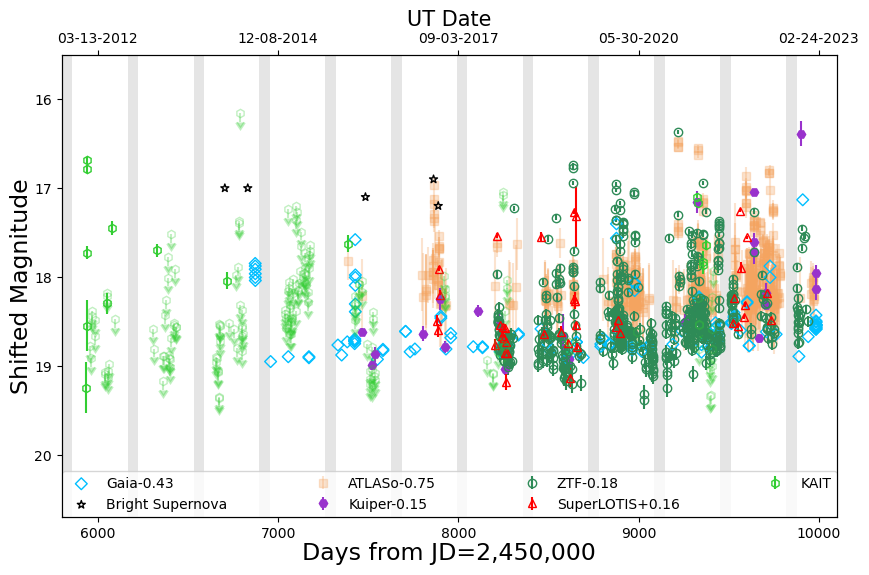}
  \label{fig:AllLCwshift}}
  
\caption{The observed light curve of AT~2016blu, mostly in $R$, but ATLAS data are in the ``orange'' ($o$) filter, {\it Gaia} data are in the $G$ band, and Bright Supernovae website data are in various filters or unfiltered. The top panel shows the light curve from 1999 to 2009, where only KAIT upper limits are available. The bottom panel shows the light curve from 2012 through February 2023. The light curve is displayed with flux adjustment for each dataset to roughly match them with one another. Vertical grey areas indicate times each year, from August 26 until October 24, when AT~2016blu is difficult to observe owing to its proximity to the Sun in the sky. Photometric observations cover at least 19 outbursts in 2012--2022. See Fig.~\ref{fig:RecentZoomed}, which zooms-in on individual eruption events.}\label{fig:AllLC}
\end{figure*}

The full light curve of AT~2016blu is shown in Fig.~\ref{fig:AllLC}; Fig.~\ref{fig:LCUL} displays the light curve from 1999 through 2009, while Fig.~\ref{fig:AllLCwshift} shows the light curve from 2012 through February 2023. 

In  Fig.~\ref{fig:LCUL} we see that the host galaxy of AT~2016blu was monitored by KAIT for $\sim 11$~yr, with no detections of a source at AT~2016blu's position. 
Based on these KAIT upper limits, the magnitude of AT~2016blu remained $\sim 19$ or fainter.  This is fainter than many of the outburst peaks seen in the 2012--2022 decade, and there are KAIT upper limits at times when we may expect an outburst based on the quasiperiodicity found below from the more recent data, indicating that AT~2016blu was not experiencing its repeated eruptions before 2009.  This may be important for understanding the origin of AT~2016blu's eruptions, and we return to this fact later.   In  Fig.~\ref{fig:LCUL}, we also see that the progenitor star detected in {\sl HST\/} images (discussed below) was safely below the KAIT upper limits around the same time period, consistent with KAIT nondetections of the quiescent progenitor star.

The more recent decade (Fig.~\ref{fig:AllLCwshift}) demonstrates much greater activity in AT~2016blu.   Super-LOTIS data are in the Johnson–Bessell photometric system. Unfiltered KAIT photometry is roughly similar to the $R$ band \citep{L03}. ZTF observations are in the ZTF $r$ filter, which is close to the SDSS $r$ filter; ZTF data are therefore scaled to Johnson-Bessell $R$ by applying a shift of $-0.18$~mag.
The {\it Gaia} $G$ band approximately corresponds to SDSS $g$, and the $o$-band ATLAS filter roughly corresponds to SDSS $r+i$. Data from the Bright Supernovae website are a mixture of various filters ($U$, Clear, and $V$). As we mentioned before, AT~2016blu is also coincident with an  H~{\sc ii} region, which can lead to flux contamination. Instead of transforming the bands, we adjust fluxes from each dataset to approximately match all datasets. The offset values are either estimated based on the flux difference between two datasets when they coincided in time, or they are determined visually to make the base quiescent magnitudes between the two datasets the same. Fig.~\ref{fig:AllLCwshift} shows the light curve after flux adjustment. All of the offsets indicated in Fig.~\ref{fig:AllLCwshift} correspond to a shift associated with a base magnitude of $\sim 18.5$. Since our goal is to search for periodicity, not derive physical properties, these shifts are merely for clarity of displaying the light curves.  

The solid vertical grey bands indicate the time interval each year, from August 26 to October 24, when AT~2016blu is difficult or impossible to observe owing to its position being too close to the Sun in the sky. It is expected that some outbursts fall within these unobservable windows or at a boundary, since the outbursts seem to repeat with a period of less than a year (see Section~\ref{sec:Priodicity} for more details). 

\begin{figure*}
 \includegraphics[width=0.5\linewidth]{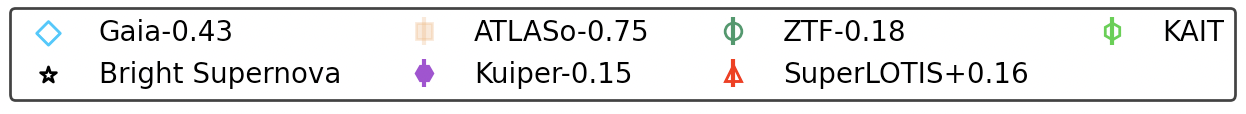}
 
\begin{tabular}{ccc}
\subfloat[]{\label{fig:RecentZoomed1}
    \includegraphics[width=0.21\linewidth]{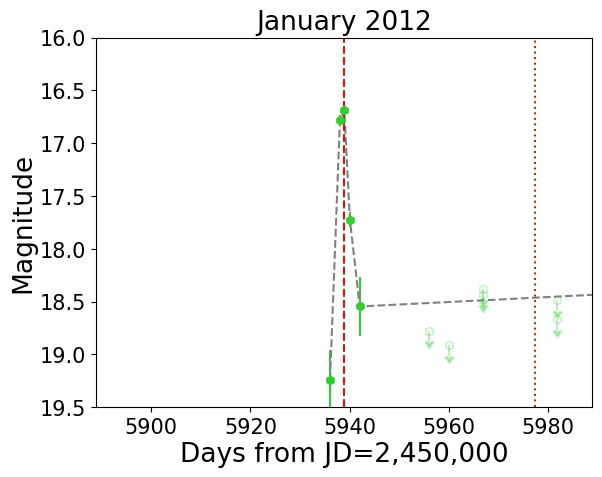}}&\vspace{-0.1 in}
    \subfloat[]{\label{fig:RecentZoomed2a}
    \includegraphics[width=0.21\linewidth]{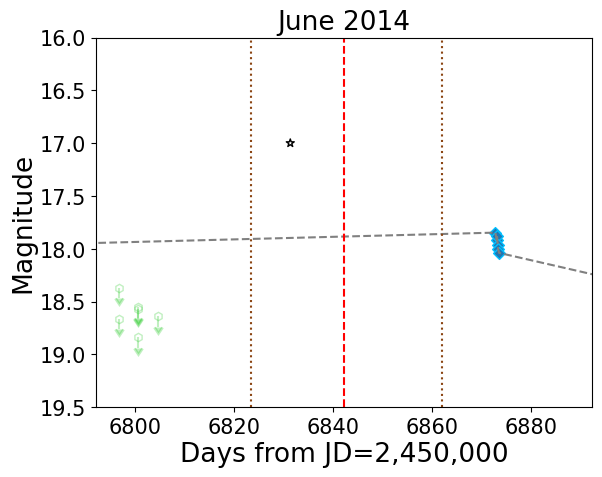}}&
    \subfloat[]{\label{fig:RecentZoomed3}
    \includegraphics[width=0.21\linewidth]{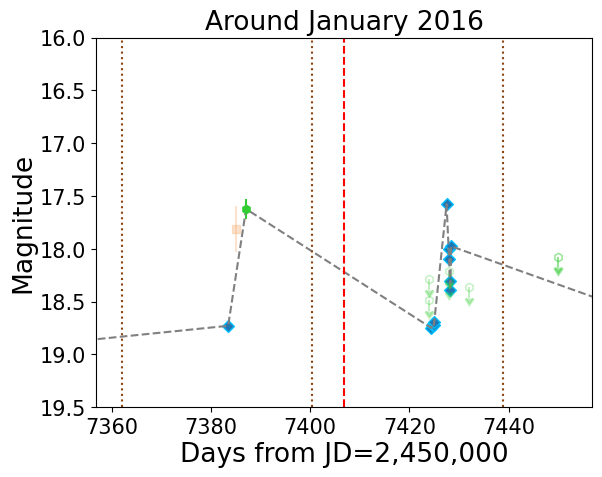}}\\
    
    \subfloat[]{\label{fig:RecentZoomed4}
    \includegraphics[width=0.21\linewidth]{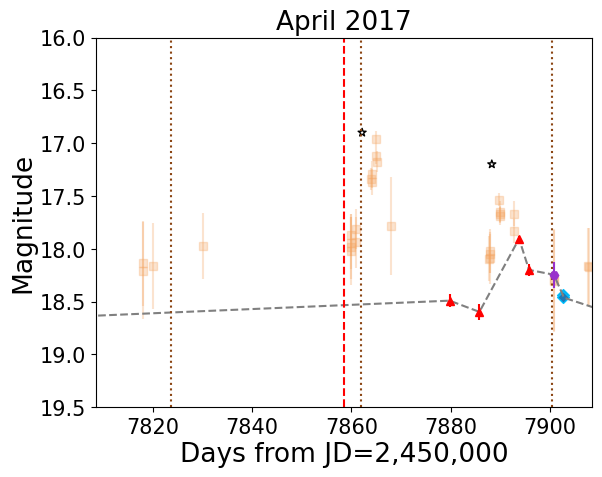}}&\vspace{-0.1 in}
    \subfloat[]{\label{fig:RecentZoomed5}
    \includegraphics[width=0.21\linewidth]{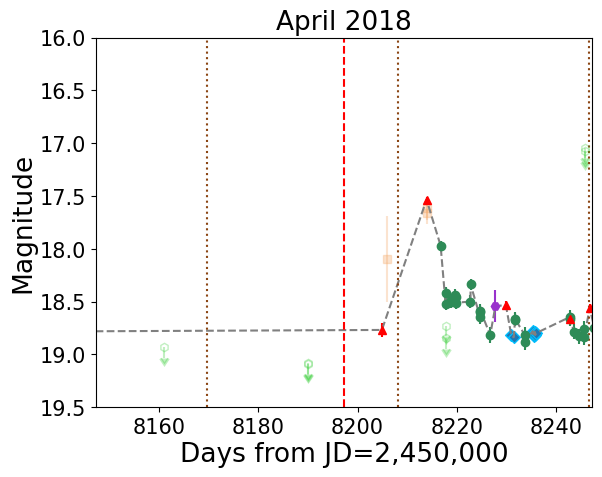}}&
    \subfloat[]{\label{fig:RecentZoomed6}
        \includegraphics[width=0.21\linewidth]{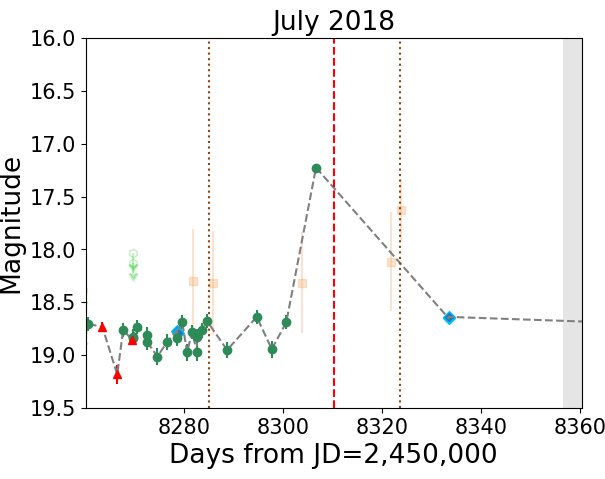}}\\
        
        \subfloat[]{\label{fig:RecentZoomed7}
     \includegraphics[width=0.21\linewidth]{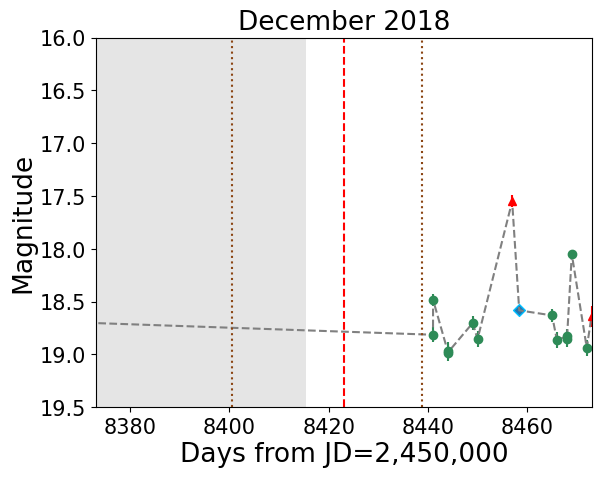}}&\vspace{-0.1 in}
     \subfloat[]{\label{fig:RecentZoomed8}
     \includegraphics[width=0.21\linewidth]{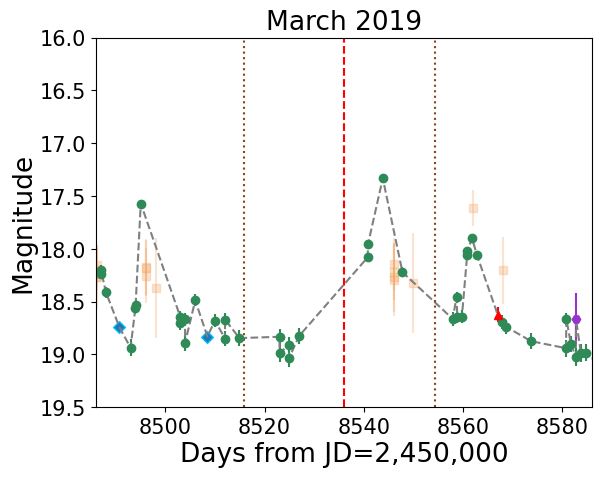}}&
     \subfloat[]{\label{fig:RecentZoomed9}
    \includegraphics[width=0.21\linewidth]{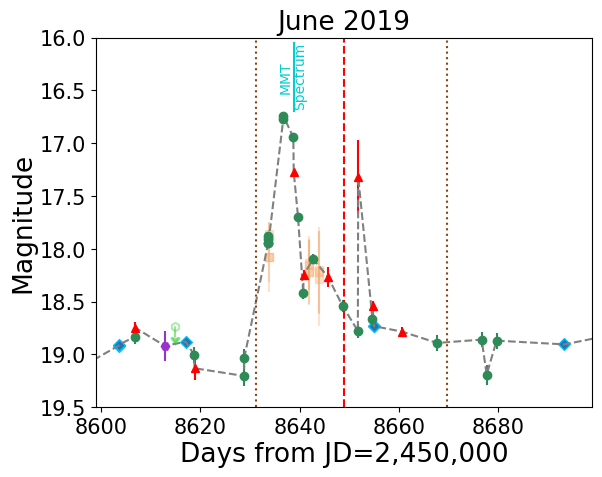}}\\
    
    \subfloat[]{\label{fig:RecentZoomed10}
    \includegraphics[width=0.21\linewidth]{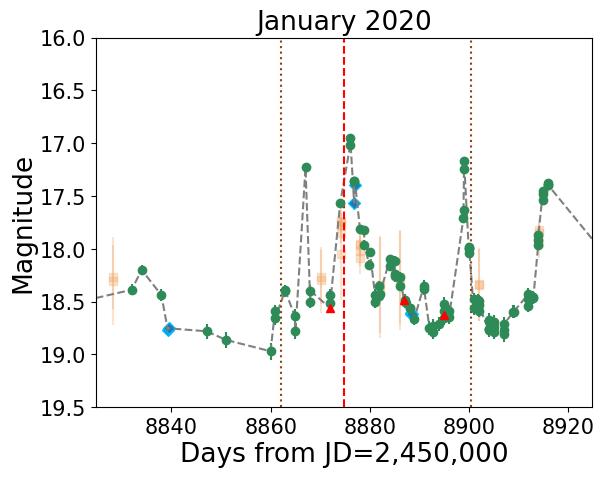}}&\vspace{-0.1 in}
    \subfloat[]{\label{fig:RecentZoomed11}
    \includegraphics[width=0.21\linewidth]{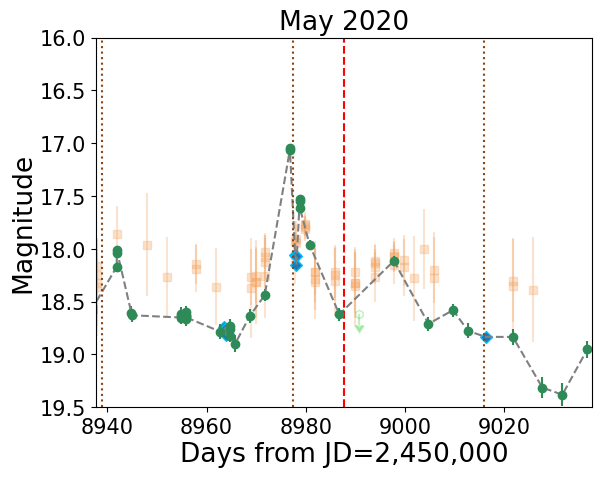}}&
    \subfloat[]{\label{fig:RecentZoomed12}
    \includegraphics[width=0.21\linewidth]{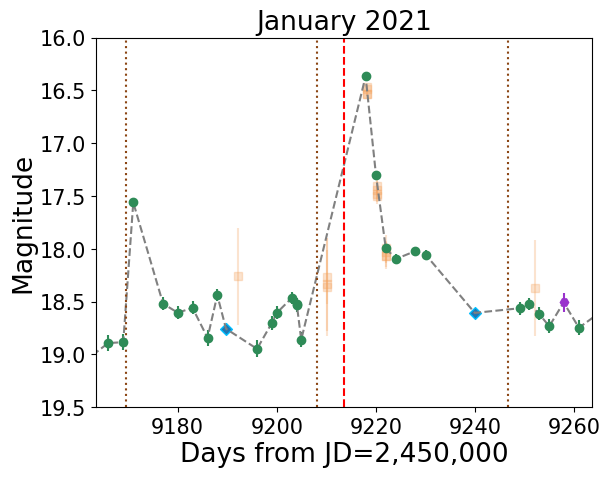}}\\
    
    \subfloat[]{\label{fig:RecentZoomed13}
    \includegraphics[width=0.21\linewidth]{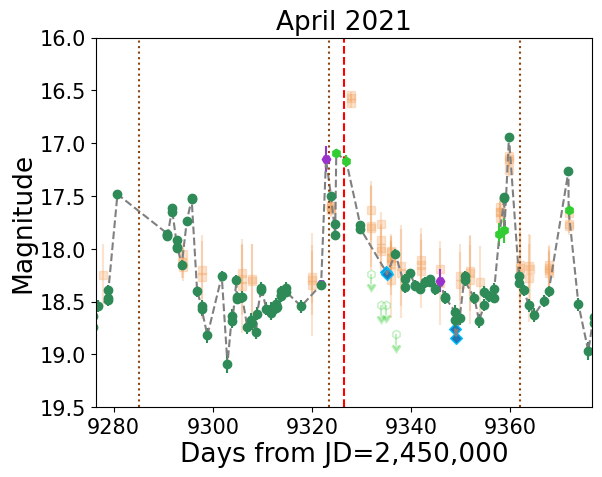}}&\vspace{-0.1 in}
    \subfloat[]{\label{fig:RecentZoomed13a}
    \includegraphics[width=0.21\linewidth]{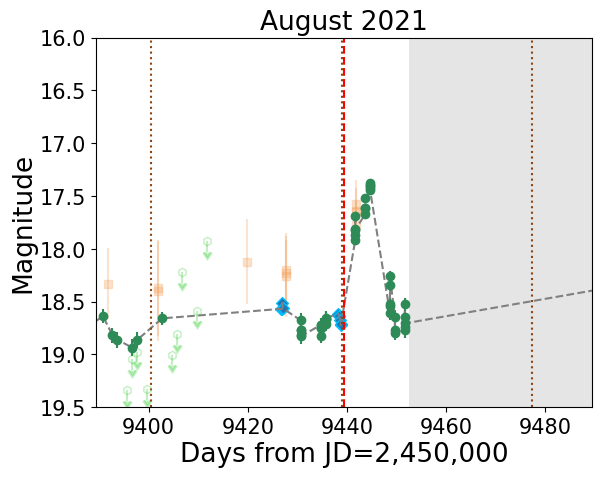}}&
      \subfloat[]{\label{fig:RecentZoomed14}
    \includegraphics[width=0.21\linewidth]{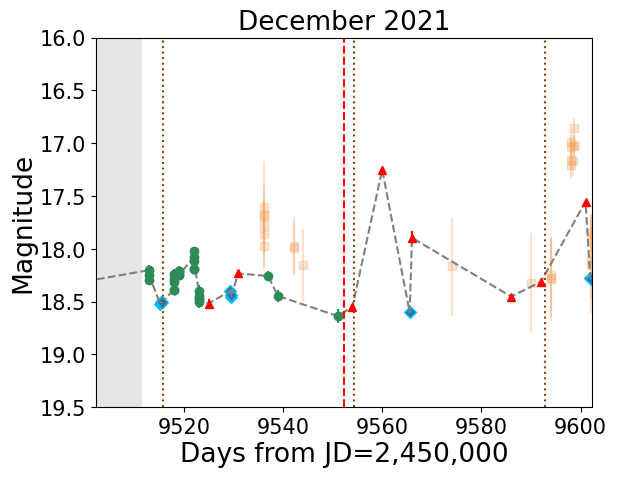}}\\
    
    \subfloat[]{\label{fig:RecentZoomed15}
    \includegraphics[width=0.21\linewidth]{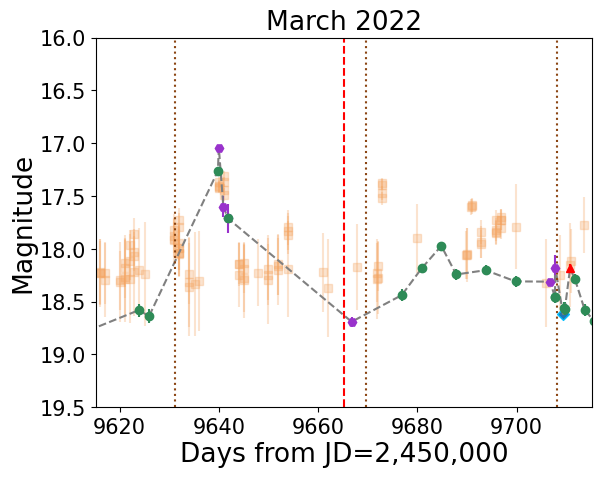}}&\vspace{-0.1 in}
    \subfloat[]{\label{fig:RecentZoomed16}
    \includegraphics[width=0.21\linewidth]{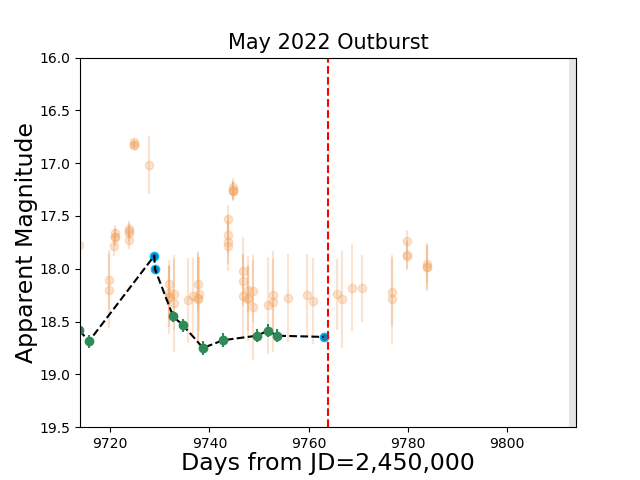}}&
\end{tabular}
\caption{Zoomed-in light curves of AT~2016blu during its multiple outbursts. The data points from all surveys, excluding ATLAS, are connected together with a dashed line. These data points are used to estimate the period. The approximate time of each outburst is given at the top of the frame. The red dashed line shows the reference epoch adopting the most likely period of $\sim 113$~d from the periodogram, tied to the first outburst in January 2012. On the other hand, the brown dotted line shows the reference epoch using the second most-dominant period in the periodogram. Some of the outbursts align better with the smaller period of 38.5~d. Quasiperiodic behaviour is evident relative to the reference epoch.}\label{fig:RecentZoomed}
\end{figure*}

Using our photometric observations, we find that AT~2016blu experienced at least 19 outbursts during the interval 2012--2022. All of AT~2016blu's known outbursts are summarised in Table~\ref{tab:summaryoutbursts}, and light curves for various individual eruptive events are shown in several panels in Fig.~\ref{fig:RecentZoomed}. We note that AT~2016blu brightened by $\sim 0.8$ mag in May 2012 and experienced a dip in October 2019. However, the data coverage during these times is insufficient to study the detailed evolution of the light curve. We define an outburst as a brightening that is more than $\sim 1$~mag, and any clustered observations with multiple peaks are considered part of the same outburst. The end of an outburst is also defined based on the quality and sampling of the data, as the outbursts are irregular. 

In the following, we describe each event in detail.  The designation for each event below corresponds to the date of the peak brightness, represented by a data point with the brightest magnitude inside the ``outburst window”. In some cases, the time of the outburst is unclear either owing to poor data coverage or multiple rapid peaks with similar magnitudes. The designation for these events is instead based on the expected time of the outburst using the detected period (based on a periodogram analysis of the light curve; see Section~\ref{sec:Priodicity}). In each plot, the red dashed line shows 
the expected time of the outburst, which is derived from $d_{\rm red} = 2,455,938.9 + 113\,n$, while the brown dotted line shows the reference epoch using the second most-dominant period in the periodogram, which is 38.5~d. Similar to Paper I, $o$-band ATLAS data are shown in the background, but they are not used to define a period because they are in a different filter and many points are close to the limiting magnitude of that telescope ($\sim 19$~mag).

{\it January 2012:} Fig.~\ref{fig:RecentZoomed1} shows the first detected outburst of AT~2016blu. On 2012 January 12, the object reached maximum brightness, with $m_R=16.7$~mag. This event is used as the reference epoch, shown as the red dashed line in Fig.~\ref{fig:RecentZoomed} and several
figures that follow.

\begin{figure}
\includegraphics[width=0.48\textwidth]{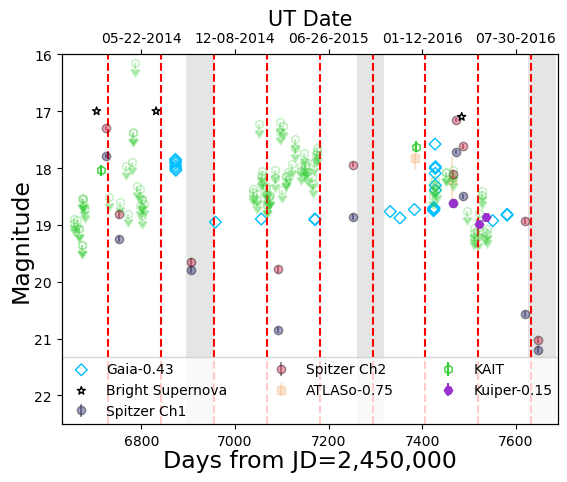}
\caption{Mid-IR light curve of AT~2016blu obtained with {\it Spitzer}, which captured two outbursts of AT~2016blu in 2014 and 2016. The Ch1 and Ch2 {\it Spitzer} data have a constant baseline flux subtracted; this flux is taken from the minimum observed flux before 2012, which was recorded in May 2004. (This baseline IR flux may include emission from the quiescent progenitor or its associated stellar population.)} 
\label{fig:Spitzer}
\end{figure}

{\it February 2014:} K. Itagaki reported an outburst discovery of AT~2016blu on February 2014. During the time of the reported discovery, no additional optical data are available to study the evolution of the light curve in detail. Fig.~\ref{fig:Spitzer} shows the mid-IR light curve of AT~2016blu, which confirms the outburst reported by K. Itagaki.

{\it June 2014:} In mid-2014, G. Cortini reported the discovery of another outburst of AT~2016blu. Shortly thereafter, the object experienced a slight decline in its brightness (see Fig.~\ref{fig:RecentZoomed2a}). Given our derived period of $\sim 113$~d (see Section~\ref{sec:Priodicity}), the time of the reported discovery is within $\sim 11$~d of an expected event.

{\it January 2016:}  Fig.~\ref{fig:RecentZoomed3} shows that around December 2015, AT~2016blu reached an apparent magnitude of 17.6. Soon thereafter in February 2016, the Gaia Photometric Alerts page reported that AT~2016blu reached a maximum brightness of 17.6~mag. During the \nth{4} outburst, AT~2016blu experienced multiple peaks with similar magnitudes. A brighter peak could have been missed if it occurred between these peaks when there was a lack of data. Therefore, the designation for this event is associated with the expected time of the outburst (red dashed line).

{\it April 2016:} In mid-2016, R. Arbour reported the discovery of another outburst of AT~2016blu \citep{A16}. Similar to previous events, the exact time of the \nth{5} outburst is uncertain owing to the lack of optical data. Nonetheless, Fig.~\ref{fig:Spitzer} shows that this outburst was also captured in mid-IR observations.
This source was named AT~2016blu by R. Arbour after the April 2016 outburst, even though it experienced multiple outbursts before 2016. Prior to this event, the source had been referred to as NGC~4559OT.

{\it April 2017:}  AT~2016blu experienced two bright phases in April and May 2017 reported in the Bright Supernovae webpage. Fig.~\ref{fig:RecentZoomed4} shows that ATLAS data reinforce the first peak brightness, while the secondary peak was covered by both ATLAS and Super-LOTIS.

{\it April 2018:} In 2018, AT~2016blu experienced its \nth{7} recorded outburst, when it reached an apparent magnitude of 17.5 (Fig.~\ref{fig:RecentZoomed5}). The observed peak of the outburst corresponds more closely to a shorter period of 38.5~d (brown dotted line) than to a longer period of $\sim 113$~d (red dashed line). Perhaps the main peak is missed in this case.

{\it July 2018:} AT~2016blu experienced its \nth{8} outburst in July 2018, when it brightened by 1.7~mag, and it reached an apparent magnitude of 17.2; see Fig.~\ref{fig:RecentZoomed6}.

{\it December 2018:} In December, AT~2016blu brightened again and reached $m_R= 17.6$~mag. The main peak might have been missed because AT~2016blu was difficult to observe prior to this time (Fig.~\ref{fig:RecentZoomed7}).

{\it March 2019:} In 2019, AT~2016blu experienced multiple fluctuations in its apparent magnitude. The March 2019 event shows spikes with similar magnitudes instead of a big spike in its brightness, as illustrated in Fig.~\ref{fig:RecentZoomed8}.

{\it June 2019:} Fig.~\ref{fig:RecentZoomed9} shows that in 2019, AT~2016blu experienced a sharp brightening by $\sim 2.5$ mag, followed by a secondary peak similar to its previous events. The first peak is more consistent with the shorter period, and the second peak is more consistent with the longer period. This suggests that the second dominant peak in the periodogram may be attributed to multiple luminosity spikes occurring in each event. Fig.~\ref{fig:spectrum} displays the MMT Blue Channel spectrum of AT~2016blu (1200 lines~mm$^{-1}$ grating) obtained
on 2019 June 4, coincident in time with AT~2016blu’s \nth{11} outburst.

{\it January 2020:} AT~2016blu experienced its \nth{12} outburst in early 2020 (see Fig.~\ref{fig:RecentZoomed10}). The January 2020 event has multiple brightness peaks with approximately similar magnitudes.  The middle peak was slightly brighter, $m_R = 17.0$~mag. 

{\it May 2020:} In May 2020, AT~2016blu reached an apparent magnitude of 17.0, see Fig.~\ref{fig:RecentZoomed11}. The observed peak aligns more with the expected time of the outburst based on the shorter period. The dip in the middle of the two main peaks is reminiscent of the dips seen during some eruptions in SN~2000ch's light curve, for which extinction from new dust formation was a suggested cause (see Paper I for more details).  Alternatively, the dips might be artificial if they are the result of the combination of the different bands from different telescopes. 

{\it January 2021:} Fig.~\ref{fig:RecentZoomed12} shows that in early 2021, AT~2016blu experienced multiple closely spaced peaks of varying brightness, somewhat similar to the March 2019 event. The second peak is brighter, where AT~2016blu brightened by 2.6~mag. 

{\it April 2021:} Again similar to previous events, AT~2016blu experienced multiple fluctuations with approximately similar magnitudes in mid-2021 (Fig.~\ref{fig:RecentZoomed13}). Although the peak observed in May 2021 is slightly brighter than the one in April, we designate the peak in April as the main event because it aligns better with the estimated time of outburst. ATLAS data show a spike in April 2021, consistent with the peak, where AT~2016blu reached $m_R=17.1$~mag.

{\it August 2021:} Later in August, AT~2016blu experienced another outburst right before it became difficult to observe (Fig.~\ref{fig:RecentZoomed13a}). During its \nth{16} outburst, AT~2016blu brightened by 1.5~mag.

{\it December 2021:} In late 2021, AT~2016blu reached $m_R=17.3$~mag. Fig.~\ref{fig:RecentZoomed14}) shows that AT~2016blu experienced two more fainter peaks subsequently, and one of them is supported by ATLAS data.

{\it March 2022:} Fig.~\ref{fig:RecentZoomed15} shows that in March 2022, AT~2016blu experienced its \nth{18} outbursts, where it brightened again and reached $m_R= 17.0$~mag. The observed peak is more consistent with the shorter period (brown dotted line) rather than the longer period (red dashed line), unless the main peak was missed.

{\it November 2022:} In late-2022, AT~2016blu reached $m_R= 16.4$~mag; see Fig.~\ref{fig:RecentZoomed16}. During this outburst, AT~2016blu brightened by 2.5~mag. Similarly to the previous event, the observed peak is more in agreement with the expected time of the outburst based on the shorter period.

\begin{table*}
\caption{Summary of AT~2016blu's 19 known outbursts from 2012 to 2022.}
\begin{adjustbox}{max width=\textwidth,center}
\renewcommand{\arraystretch}{1.}
\begin{tabular}{ccccccc} 
\hline
Outburst &Peak& Expected Outburst Time&Approx. Amplitude & Approx. Duration & Ref. &  \\
$\#$ &(JD$-$2,450,000) & (JD$-$2,450,000)& (mag) & (d) &  \\
\hline
1& 5938.9& 5938.9 &2.6&6.0& LOSS/This work\\
2&6704.2&6729.4&...&...& Itagaki$^*$\\
3&6831.4&6842.3&...&...& Cortini$^*$ \& This work\\
4&7427.5&7406.9&1.2&45.0&This work\\
5&7484.4&7519.8&...&...& Arbour$^*$\\
6&7862.2&7858.6&...&...& Itagaki, Arbour$^*$ \& This work\\ 
7&8213.9&8197.3&1.3&$>$31.1&This work\\ 
8&8306.7&8310.2&1.7&35.7&This work\\
9&8457.0&8423.2&1.4&$>$31.0&This work\\
10&8543.8&8536.1&1.7&75.8&This work\\
11&8636.7&8649.0&2.5&26.3&This work\\
12&8876.0&8874.8&2.0&84.9&This work\\
13&8976.8&8987.8&1.9&39.0&This work\\
14&9217.9&9213.6&2.6&70.9&This work\\

15&9324.9&9326.5&2.1&96.8&This work\\
16&9444.6&9439.4&1.5&24.9&This work\\
17&9560.0&9552.4&1.5&62.7&This work\\

18&9640.0&9665.3&1.6&42.9&This work\\
19&9903.0&9891.1&2.5&64.1&This work\\

\hline
\label{tab:summaryoutbursts}
\end{tabular}
\end{adjustbox}
$^*${\tt \url{https://www.rochesterastronomy.org/sn2012/lbvn4559.html}}
\end{table*}

\section{Wavelength Dependence}\label{sec:Wavelength} 
\subsection{Variability in the Mid-Infrared}\label{sec:spitzer} 
As we mentioned in the previous section, AT~2016blu's two outbursts in 2014 and 2016 were also observed by {\it Spitzer}. Fig.~\ref{fig:Spitzer} shows the mid-IR and optical light curve of AT~2016blu. These outbursts were also supported by optical observations made by amateur astronomers as discussed in Section~\ref{sec:LC}. An IR source coincident with AT~2016blu was also detected back in 2004 (not shown in Fig.~\ref{fig:Spitzer}; see Table~\ref{tab:spitzer}), but it is not clear whether AT~2016blu was in outburst or quiescence during that time; moreover, that IR flux may include light from neighbouring stars or the nearby H~{\sc ii} region, as discussed below. 

Given that the quiescent {\it Spitzer} flux may be contaminated by some light from the underlying stellar population, and since we are most interested in the change in IR excess emission during outburst, in Fig.~\ref{fig:Spitzer} we show the {\it Spitzer} data after subtracting the minimum observed flux before 2012, which was recorded in May 2004. It is clear that after this subtraction, there is significant excess IR flux that is due to transient emission from eruptions of AT~2016blu. The observed excess IR emission could be due to (1) blackbody radiation from the photosphere of AT~2016blu itself, (2) newly formed dust in material ejected by AT~2016blu, or (3) pre-existing CSM dust ejected previously by AT~2016blu that is heated by the rising luminosity of the outburst. It is possible to determine whether IR emission is caused by dust emission or originates from enhanced radiation from the photosphere via {\it Spitzer} colours. The colour estimates in Table~\ref{tab:spitzer} range from 0.14 to 1.63~mag, which are redder than the expected {\it Spitzer} colour in these bands for the Rayleigh-Jeans tail of the transient's photosphere. The expected colour for a star with a temperature of higher than 7000~K (see Section~\ref{sec:Precursor} for more details regarding the estimated temperature) is 0--0.1~mag or bluer, as shown in Fig.~7 of \citet{B10}. The detection of an IR excess suggests the presence of dust in the vicinity of AT~2016blu.  Owing to a lack of sufficient optical observations during the IR peaks, we are unable to determine whether the dust is newly formed or pre-existing.

\subsection{Visual-Wavelength Colour Evolution}\label{sec:color} 
\begin{figure*}
\subfloat[]{
 \hspace*{-0.03\textwidth} \includegraphics[width=0.8\textwidth]{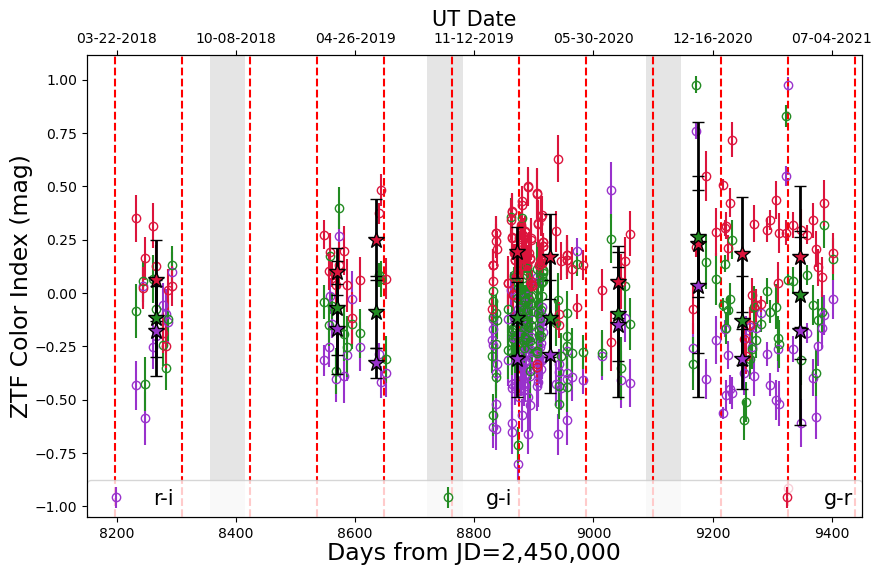}
  \label{fig:Color}}\\
\subfloat[]{
  \includegraphics[width=0.8\textwidth]{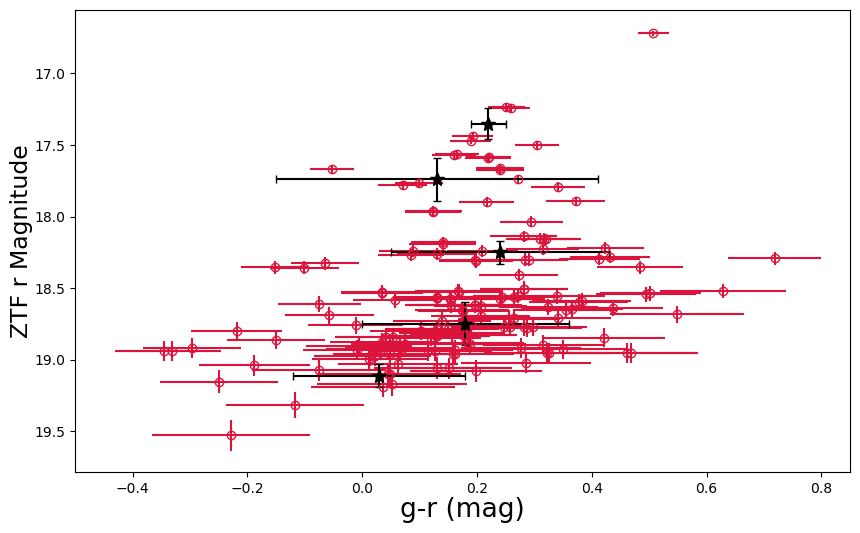}\hspace*{+0.2 in}
  \label{fig:Colormag}}
  \caption{Colour evolution of AT~2016blu using only ZTF Sloan filters, with no magnitude shift.  The top panel shows the colour evolution over time in 
 $r-i$ (purple), $g-i$ (green), and $g-r$ (red). The points marked with a star symbol represent the average colour obtained by binning the colour light curve. The bottom panel shows the ZTF $r$-band magnitude as a function of $g-r$ colour. Black points represent the average colour and $r$ magnitude for each magnitude bin. There is no significant evidence of a change in colour over time for AT~2016blu.}\label{fig:ColorALL}
\end{figure*}

Fig.~\ref{fig:ColorALL} shows the colour evolution of AT~2016blu based solely on ZTF data without any magnitude shift. As multiband observations were not taken simultaneously, we interpolate the dataset with a larger number of data points ($r$ and $g$ bands), in order to match the sampling size of the smaller dataset ($i$ band). This allowed us to approximate the $r$- and $g$-band data at the same time as the $i$-band data. When we interpolate the data, we also need to account for the uncertainty in the measurements. To do this, we assign the magnitude uncertainty of the interpolated data to be the same as the magnitude uncertainty of the closest magnitude in that specific band. In other words, we look for the closest measurement in the same band and use its magnitude uncertainty as the uncertainty for the interpolated data. The colour uncertainty corresponds to the uncertainties in each band added together in quadrature. 

Fig.~\ref{fig:Color} shows that the colour of AT~2016blu does not exhibit a consistent pattern with respect to times of outburst, with relatively small colour changes of $\la 0.5$~mag in any individual colour. Fig.~\ref{fig:Colormag} shows $r$ magnitude as a function of $g-r$ colour, which is meant to determine if the colour changes systematically with brightness level.  In order to assess whether there is a significant change in colour over time, the data have been grouped into bins of size  0.5~mag (indicated by black points). The difference in colour is smaller than the deviation and the data are therefore in agreement with the same colour over time. Overall, this indicates that any colour shift is small ($\la 0.25$~mag) compared to the amplitude range of eruptions (1--2~mag).  While traditional LBVs are expected to become much cooler as they brighten, a subset of LBV-like transients including SN~2009ip's progenitor, SN~2000ch, and M33 MCA-1B appear to remain hot at their peak brightness level \citep[Paper I;][]{S10,S11,S20}.

Next, we investigate the possible presence of periodic behaviour in the light curve of AT~2016blu, utilising the Lomb-Scargle periodogram. This statistical tool is commonly used for detecting periodic signals in observations that are not evenly spaced. As in the case of SN~2000ch (Paper I), finding periodic behaviour could give a hint of the underlying mechanism that triggers these outbursts.

\section{Search for Periodicity}\label{sec:Priodicity} 
In this section, we use the methodology presented in Paper I to identify possible periodic behaviour in the light curve of AT~2016blu. First, we use a Lomb-Scargle periodogram, and then we fold the light curve to analyse its period variation. 

\begin{figure}
\includegraphics[width=0.45\textwidth]{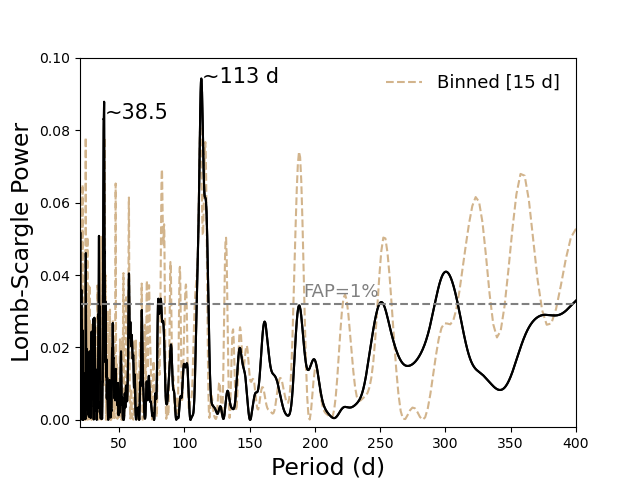}
\caption{The Lomb-Scargle periodogram of AT~2016blu's light curve using optical photometry data from various facilities spanning from 2012 through February 2023 (excluding upper limits and ATLAS data). We find that the outbursts seem to repeat with a most likely period of 112.9~d. Two smaller peaks with lower amplitudes can also be observed at $\sim 35$~d and $\sim 38$~d. These peaks are likely to be associated with multiple spikes occurring either before or after each outburst. The 1\% FAP level is indicated by a horizontal dashed line. Additionally, we generated a periodogram of the binned light curve (brown dashed line), which shows a similar optimal period when the bin size is less than $\sim 15$~d. However, larger bin sizes result in a periodogram without a single strong peak because the bright peaks get removed during binning.} \label{fig:periodogram}
\end{figure}

Fig.~\ref{fig:periodogram} shows the Lomb-Scargle periodogram \citep{L76, S82, v18} of AT~2016blu's light curve, for possible periods between 20~d and 400~d. We find that  AT~2016blu's eruptive outbursts repeat with a best period of 112.9~d. The false-alarm probability (FAP) of this peak is very small, of the order of $10^{-14}$, meaning that this peak is statistically significant.  The horizontal dashed line in Fig.~\ref{fig:periodogram} shows the 1\% FAP level. Based on the duration of the outbursts, the uncertainty in the period should be $\sim 2$~d.

Two smaller-amplitude peaks can also be seen around 35~d and 38~d. As we discussed in Section~\ref{sec:LC}, AT~2016blu experienced multiple fluctuations in its apparent magnitude. We find the separation between closely-spaced peaks to be
20--50~d, suggesting that the two low-period peaks in the periodogram might be due to multiple rapid spikes in luminosity. In Section~\ref{sec:LC}, we demonstrated that some of the luminosity peaks are more closely aligned with the shorter period of 38.5~d (see dashed brown lines in Fig.~\ref{fig:RecentZoomed}).

We conduct additional analyses by binning the light curve using various bin sizes and performing a search on the binned light curve to verify the robustness of our primary findings. Our results show that bin sizes less than $\sim 15$~d result in a similar optimal period as our primary analysis. However, larger bin sizes do not reveal a single strong peak in the periodogram. This outcome is expected because a bin size of more than 15 days becomes larger than the peak duration of most of the outbursts, and therefore begins to dilute or erase the signal.

\begin{figure}\begin{center}
\includegraphics[width=0.33\textwidth]{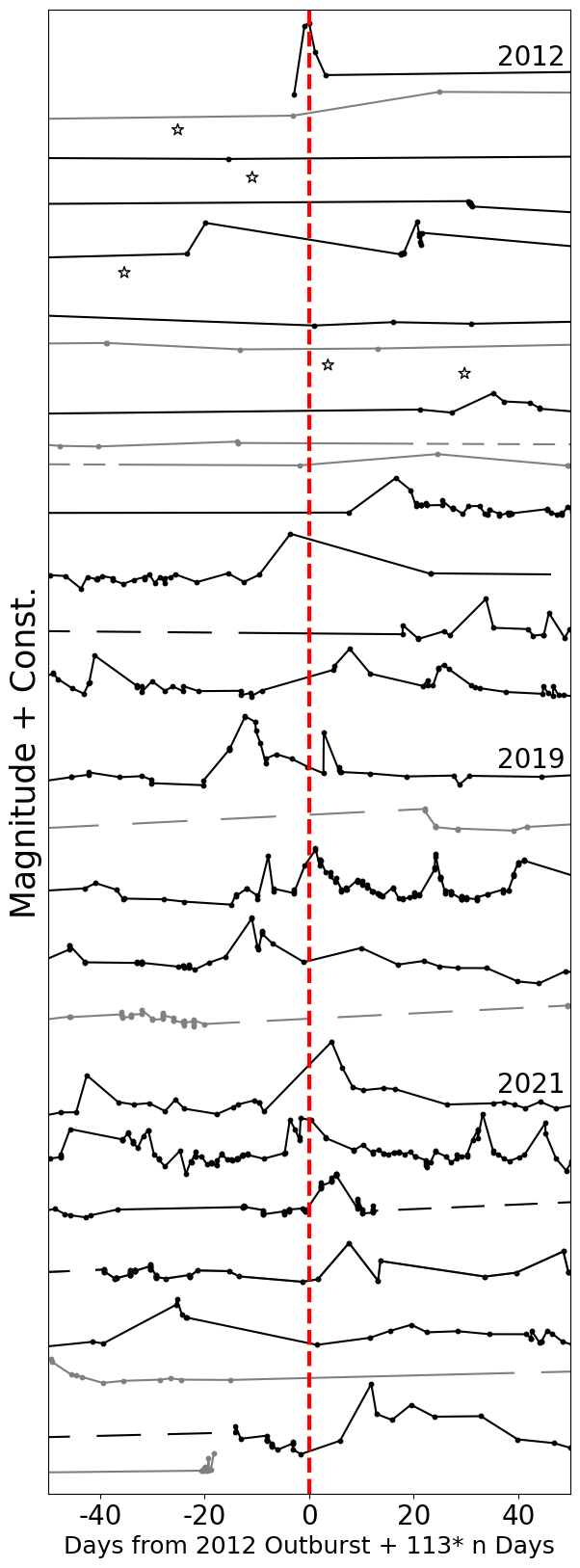} \end{center}
\caption{Folded light curve of AT~2016blu  using a period of $\sim 113$~d. ATLAS data and points with upper limits are not shown to reduce clutter in the figure. The red vertical dashed line identifies the reference epoch. All cycles with good data coverage are shown. Dashed grey/black lines indicate the time interval when AT~2016blu was difficult to observe during each cycle. Black lines represent the evolution of AT~2016blu during 19 major outbursts. Quasiperiodic behaviour can be seen from the light curve of AT~2016blu, since peaks tend to occur near the red dashed line (or on either side of it when there are multiple peaks).}
\label{fig:FoldedLC}
\end{figure}

We then carry out the period analysis by folding the light curve with the detected period of $\sim 113$~d. Fig.~\ref{fig:FoldedLC} shows the folded light curve of AT~2016blu, which is like a two-dimensional observed minus calculated (O$-$C) plot. All the data are time-shifted by $n \times 113$ days ($n$ is from 0 to 26). Cycles with poor data coverage are excluded, and dashed lines show the months (August--October) when AT~2016blu was difficult to observe. Black lines indicate the evolution of the variable star during the 19 major outbursts. The red vertical dashed line
shows the reference epoch, which is derived
from the first outburst in January 2012. Similar to SN~2000ch's light curve (Paper I), AT~2016blu's light curve exhibits quasiperiodic variation, but AT~2016blu's outbursts are more irregular. While many individual outbursts align well with the predicted time, some --- especially cases where there are multiple outbursts --- seem to straddle the expected time or are offset slightly. Several outbursts may have been missed for different reasons: (1) outbursts coincident in time with when AT~2016blu could not be observed owing to its position relative to the Sun, and (2) poor data coverage at the time of the outburst. It is unlikely that the outbursts look dimmer owing to dust formation close to the star. If faint points were intrinsically bright but appeared faint because of extinction, they would also be systematically reddened and would therefore be located towards the lower right in Fig.~\ref{fig:Colormag}, which is currently empty.

\begin{figure*}
\subfloat[]{
  \includegraphics[width=0.75\textwidth]{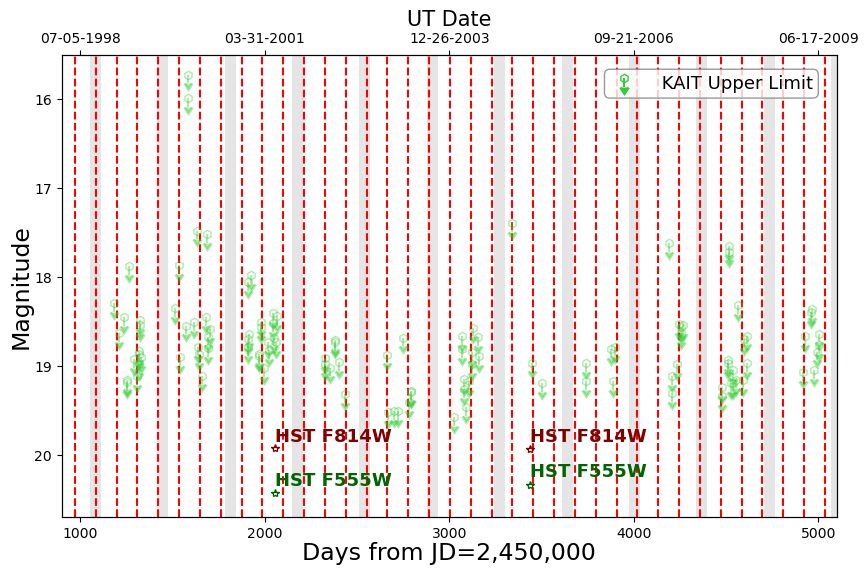}
  \label{fig:LCULw112}}\\
\subfloat[]{
  \includegraphics[width=0.75\textwidth]{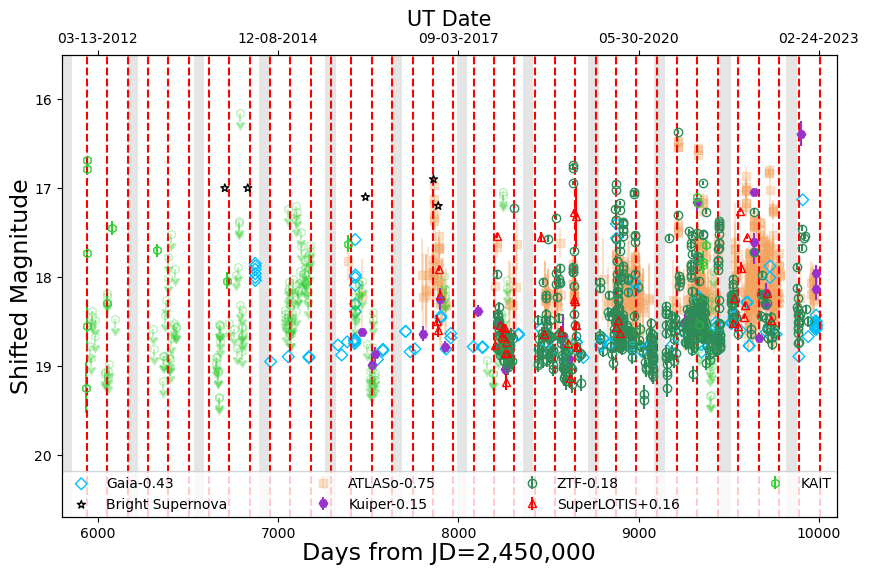}
  \label{fig:AllLCwshiftw112}}  
\caption{Same as Fig.~\ref{fig:AllLC}, but with dashed red lines to indicate the reference epoch using a period of $\sim 113$~d. The detected period matches with most of the outbursts, suggesting that the period is not changing much over time.}\label{fig:LCperiod}
\end{figure*}

Fig.~\ref{fig:LCperiod} displays the full light curve again with calculated times of events using a period of $\sim 113$~d (red vertical lines). Fig.~\ref{fig:LCULw112} shows the light curve over a period of $\sim 11$~yr where only upper limits are available. Some upper limits overlap with the predicted times of the outbursts. This implies that AT~2016blu did not experience outbursts prior to 2009. However, since there is a lack of data between mid-2009 and 2012 (the KAIT data were lost, unfortunately, because two storage disks malfunctioned), we cannot determine whether the outburst began before 2012 or the 2012 event was the first outburst. Fig.~\ref{fig:AllLCwshiftw112} shows the light curve from 2012, and the period of $\sim 113$~d matches most of the outbursts, considering that some of the main brightness peaks might be missed as we described above. This suggests that
the period is not changing significantly over time. There are KAIT upper limits that overlap with some predicted times of the outbursts after 2012. However, it is challenging to determine whether AT~2016blu was undergoing an outburst during those times because brightness peaks are narrow and may occur just before or after the expected time.

\begin{figure}
 \includegraphics[width=0.45\textwidth]{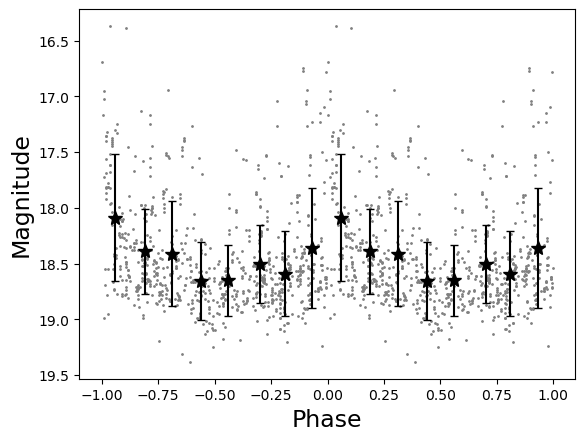}
\caption{Phase diagram of AT~2016blu using the period of $\sim 113$~d. Black points show average magnitudes after binning the folded light curve. AT~2016blu does tend to get brighter around phase zero.}\label{fig:phasediagram}
\end{figure}

Fig.~\ref{fig:phasediagram} displays the phase diagram of AT~2016blu, where all the cycles are folded on top of each other. The phase diagram shows two complete cycles, and to better see the behaviour of AT~2016blu, a folded light curve is binned. Black points in Fig.~\ref{fig:phasediagram} indicate average magnitudes after binning the folded light curve. AT~2016blu seems to get brighter around phase zero, which is set using the first outburst in January 2012.

Above we demonstrated that AT~2016blu experienced multiple outbursts over the past $\sim 11$~yr. Using the Lomb-Scargle periodogram and folded light curve, we showed that the variations are quasiperiodic. Similar to the case of SN~2000ch (Paper I), the quasiperiodic variability seen in the light curve has important implications for the cause of the outbursts. In the next section, we analyse {\sl HST\/} observations to constrain the approximate initial mass and luminosity of AT~2016blu in its presumably quiescent state before the eruptive variability began. Next, in Section~\ref{sec:binary}, we discuss an underlying mechanism that might trigger AT~2016blu's outbursts, and we note how this mechanism might account for both similarities and differences between AT~2016blu and SN~2000ch.

\section{The Progenitor}
\label{sec:Precursor}
As mentioned in Section~\ref{sec:LC}, past imaging from LOSS indicates no detection of eruptive variability in the period 1999 to 2009, despite frequent observations of the host galaxy.  This implies that the progenitor star was not in an eruptive state prior to 2009 (or possibly prior to the first eruption detected in 2012).  In that case, the {\sl HST\/} images obtained in 2001 and 2005 allow us to constrain the properties of the relatively quiescent progenitor star.

\subsection{Constraining the Properties of the Progenitor}

\begin{figure}
 \includegraphics[width=0.45\textwidth]{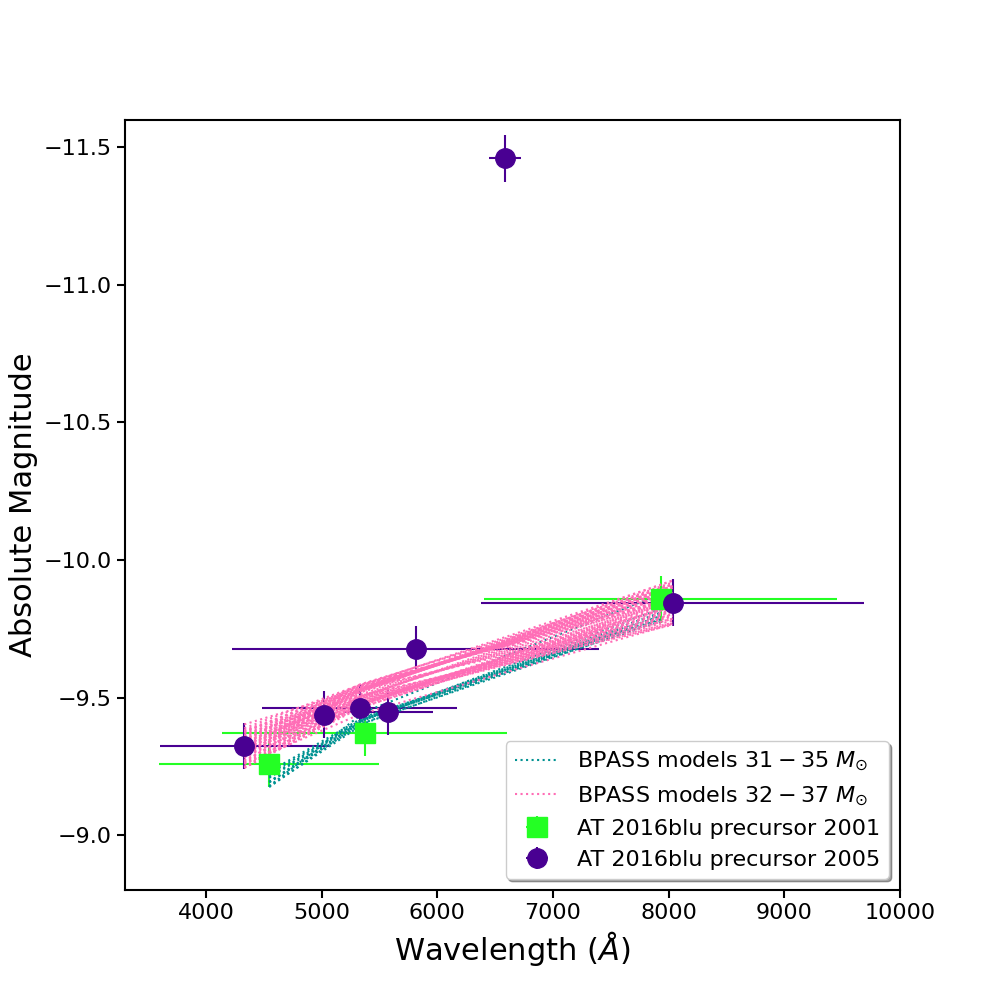}
\caption{The SED of the AT~2016blu progenitor, established from the pre-outburst {\sl HST\/} observations from 2001 and 2005 (see text). For comparison we also show BPASS single-star model SEDs, with ZAMS masses ($M_{\rm ZAMS}$) as indicated in the legend.}\label{fig:precursor_sed}
\end{figure}

Using {\sl HST\/} observations described in Section~\ref{sec:HSTobs},
we built a spectral energy distribution (SED) from the extracted photometry in all of the bands, corrected for the assumed distance to the host galaxy NGC~4559 of 8.91 ($\pm 0.19$ statistical, $\pm 0.29$ systematic) Mpc \citep{Mc17} and for foreground reddening.  
We show the resulting SED in Fig.~\ref{fig:precursor_sed}. The uncertainties in the data points are driven mostly by the uncertainty in the distance.

The SEDs constructed from the 2001 and 2005 data tend to agree, implying that the star evolved little during that time period.  The large excess in the F658N band indicates strong H$\alpha$ emission, which also affected the broad F606W band as well. Note that we assumed the host reddening to be small, equivalent to the foreground reddening, and did not apply any correction for possible circumstellar reddening (we return to this below), so values derived here are lower limits to the luminosity, temperature, and mass.

We then compared the photometry in the broad continuum bands from 2001 (F450W, F555W, F814W) and 2005 (F435W, F555W, F814W) to model SEDs from Binary Population and Spectral Synthesis (BPASS) v2.2.2 \citep{Stanway2018}. The objective was to determine approximate values of effective temperature, $T_{\rm eff}$, and bolometric luminosity, $L_{\rm bol}$, matching the intrinsic colours of the progenitor. Over the range of BPASS single-star models (with ZAMS masses $M_{\rm ZAMS}=31$--35~M$_{\odot}$ for the 2001 data and a similar 32--37~M$_{\odot}$ for the 2005 data) that agreed with the colours, these corresponded to $T_{\rm eff}=6745^{+302}_{-288}$~K and $\log (L_{\rm bol}/{\rm L}_{\odot})=5.661^{+0.023}_{-0.031}$.

\subsection{Temperature Discrepancy and Possible Solutions}

We note that a temperature of $\sim 7000$~K is at odds with the He~{\sc i} line emission observed in the spectrum (see Section~\ref{sec:Spectrum}), which would normally indicate a much higher temperature $\ga 20,000$~K. Possible solutions to this discrepancy are that (1)  AT~2016blu actually has a higher effective temperature of $\sim 20,000$~K, but appears cooler in photometry because of reddening from additional circumstellar dust (this would also require a higher implied luminosity and initial mass than that derived above),  (2) a hot companion star could assist in ionising the wind or circumbinary material, or (3) shock excitation in wind collisions or colliding CSM shells could excite the He~{\sc i}.  We defer this question to a future in-depth study of the spectral evolution of AT~2016blu; for now, we remain cognisant of the possibility that the temperature, luminosity, and implied initial mass are significantly higher than derived from the {\sl HST\/} SED, and we regard the values quoted above as lower limits.

Even without any correction for possible circumstellar dust, the lower limits on the mass and luminosity have important implications for AT~2016blu.  Fig.~\ref{fig:precursor_hrd} shows the estimated location of AT~2016blu's progenitor on the Hertzsprung-Russell (HR) diagram.  The black circle notes the estimate corrected only for line-of-sight reddening.  As noted above, this location was matched well by a BPASS model track for a single star with an initial mass of 33~M$_{\odot}$ (see Fig.~\ref{fig:precursor_hrd}).  In this particular model, AT~2016blu would be on a blue loop, requiring that it would have had a much larger radius in a previous red supergiant phase.  This might be problematic if it is in an interacting binary system, because that red supergiant would be much larger than the current orbit.  

Alternatively, the position of the AT 2016blu’s progenitor in
the HR diagram can be roughly matched using a BPASS model designed for a single star with an initial mass of 37~M$_{\odot}$ (shown with an orange solid curve). In this model, the progenitor is still evolving across the HR diagram, approaching the red supergiant phase for the first time.

Finally, the location of AT 2016blu’s progenitor on the HR diagram can also be matched well with a BPASS model for an initially 40~M$_{\odot}$ primary star in a binary system (solid red curve in Fig.~\ref{fig:precursor_hrd}). However, in the next section, we discuss the reason why none of these models provides a satisfying explanation for the progenitor of AT~2016blu.

\smallskip\smallskip

\subsection{Comparison with LBVs and Massive Stars in the Region}

\begin{figure}
 \includegraphics[width=0.47\textwidth]{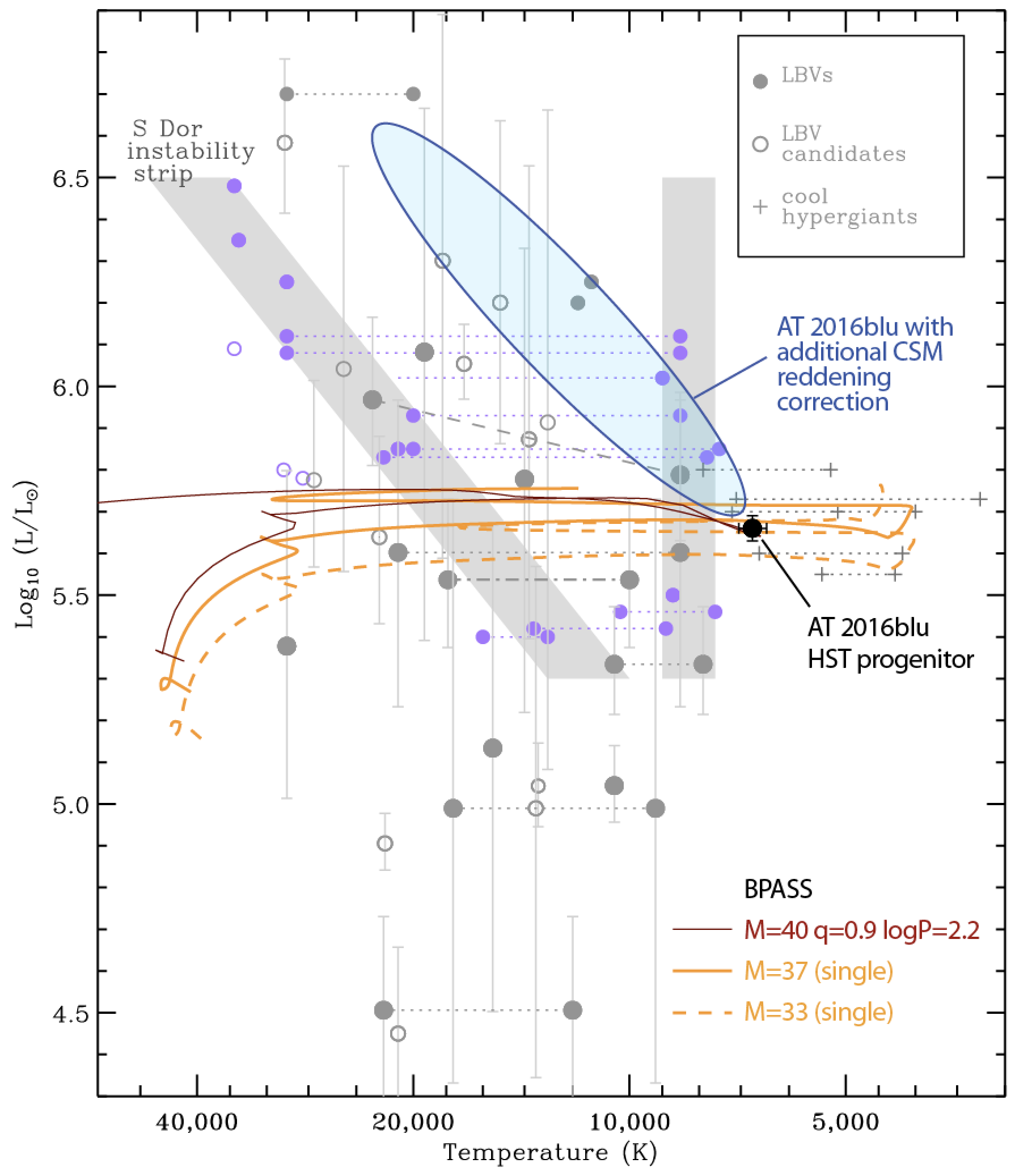}
\caption{HR diagram showing the location of the 2001--2005 progenitor detected by {\sl HST\/}, marked with a solid black circle.  For comparison, we also show values for LBVs and LBV candidates in the Milky Way (grey circles) and in the Magellanic Clouds (violet circles)  taken from \citet{smith19gaia}, as well as yellow hypergiants (grey ``plus signs") from Smith et al. 2004 (see those papers for additional details).  Also for comparison, we plot BPASS stellar evolution models for a single star with $M_{\rm ZAMS} = 33$~M$_{\odot}$ (orange dashed track), $M_{\rm ZAMS} = 37$~M$_{\odot}$ (orange solid track), and a primary with $M_{\rm ZAMS} = 40$~M$_{\odot}$ in a binary with a mass ratio $q=0.9$ and an orbital period of $\sim 150$~days (dark-red solid track).  The blue ellipse shows an approximate range of locations for the progenitor of AT~2016blu if it were to be corrected for additional extinction from an unknown amount of CSM (the black point is only corrected for foreground ISM extinction).  While we do not know the amount of CSM dust, additional reddening is motivated by the presence of He~{\sc i} emission, which likely requires $T_{\rm eff}$ significantly hotter than 7000~K (see text). }\label{fig:precursor_hrd}
\end{figure}

For comparison with AT 2016blu, Fig.~\ref{fig:precursor_hrd} also shows LBVs and yellow hypergiants in the Milky Way and Local Group.  We see that AT~2019blu lands amid the yellow hypergiants, although very close to LBVs in their cooler eruptive state indicated by the vertical grey bar.  Any additional correction for circumstellar dust would move the star's position somewhere to its upper left, indicated by the blue ellipse, and thus would locate it amid the modulations experienced by the more luminous group of classical LBVs, making it too hot to be a yellow hypergiant.  As noted earlier, the presence of He~{\sc i} emission in the spectrum gives reason to suspect that there may be additional circumstellar reddening.  No yellow hypergiant is known to show He~{\sc i} emission in its spectrum.

As previously mentioned, AT~2016blu stands out as the brightest star within the $\sim 175$~pc radius shown in Fig.~\ref{fig:images2}.  There are several other blue and red supergiants in this region, which \citet{Soria2005} estimated to have initial masses of 10--15~M$_{\odot}$ and ages of $\sim 20$~Myr, as noted earlier.  If AT~2016blu's progenitor is interpreted as a single star with an initial mass of 33~M$_{\odot}$ (and therefore an age of around 5--6~Myr), then its age is incompatible with all the neighbouring massive stars.  If it is a 40~M$_{\odot}$ primary star in a binary system (4--5~Myr), or if there is any additional circumstellar reddening, then this age discrepancy is even worse.  

The progenitor of AT~2016blu was therefore clearly a massive blue straggler.  This is something that the progenitor of AT~2016blu has in common with LBVs as well, with most LBVs being isolated from O-type stars \citep{ST15,A17}.  The rare LBVs that are found in clusters or associations with other massive stars, such as $\eta$~Car, W243 in Wd1, R127, and S~Dor, tend to stand out as far more luminous and often seem much younger than their siblings \citep{ST15,A17,A20,smith16,smith19,B21}.  Since AT~2016blu's progenitor was a massive blue straggler, it likely evolved as an interacting binary.  As such, the BPASS model tracks included in Fig.~\ref{fig:precursor_hrd} are probably not good representations of its past evolution, although they can match its current location on the HR diagram. (Even the binary BPASS track for a 40~M$_{\odot}$ primary shown in Fig.~\ref{fig:precursor_hrd} is probably not applicable to AT~2016blu, since it is a mass donor in the system, with an age that is too young.)  Instead, it is likely that the progenitor has evolved as a mass gainer in a binary or is the product of a stellar merger.

Given its location near LBVs on the HR diagram and its blue-straggler status that is shared by LBVs, we consider it likely that AT~2016blu's progenitor is indeed an LBV.  This, combined with the quasiperiodicity of its eruptions, motivates a scenario wherein  its repeating outbursts are caused by binary interaction, and where one of the stars in the system is an LBV, as discussed next.

\smallskip\smallskip

\section{LBV-Modulated Binary Interaction} \label{sec:binary}
The light-curve analysis of AT~2016blu shows periodic outbursts with irregular light curves, similar to those of SN~2000ch. In \citet{AA22}, we proposed that SN~2000ch's outbursts resemble photometric properties similar to the rapid brightening and fading observed in pre-SN eruptions of SN~2009ip \citep{S10,P13}, and $\eta$~Car's periastron encounters leading to its Great Eruption \citep{sf11,SN11}. Therefore, we proposed that SN~2000ch's outbursts occur around times of periastron in an eccentric massive binary system, where the irregularity may arise because the interaction is modulated by erratic LBV-like instability in one of the two stars. The intrinsic variability of the LBV may not look the same as in normal LBVs because the interacting companion star can lead to mass loss and hinder the expansion of the LBV, or because the intrinsic decade-long LBV brightness variations are mild compared to the more sudden and violent outbursts discussed here. 

As we discussed in Section~\ref{sec:LC}, similar to recent observations of SN~2000ch, AT~2016blu 
experienced an outburst every 3--4 months in the past decade. A regular periodicity of $\sim 113$~d implies that there are two closely interacting stars, and a binary orbit governs the repeating events. AT~2016blu's light curve also exhibits a wide variety of peak luminosity, duration, and shape. Some events exhibit multiple rises and falls. As demonstrated in Section~\ref{sec:LC}, the luminosity spikes occasionally occur before or after the anticipated time of outburst. 

The quasiperiodic variability suggests
that the binary orbit is eccentric (in other words, if the orbit were circular, then  the level of interaction would be relatively constant and would not show outburst periodicity). When the separation between the two stars decreases at periastron, the interaction between the two stars --- whatever the exact nature of that interaction may be --- should be
stronger.

The specific type of binary ``interaction'' that is responsible for the brightening may arise from a variety of physical phenomena, including strong wind collisions \citep{pc02,okazaki08,K21}, collision between the companion star and CSM (a shell or disk; \citealt{S18}), grazing stellar collisions \citep{SN11}, unsteady mass transfer and accretion \citep{S04, K09,K10}, etc. (this was discussed in detail in Paper I).  A discussion of which of these may be responsible for the interaction in this case is postponed to a later time when we analyse spectra of AT~2016blu. 

If the CSM is significantly clumpy or if accretion is unsteady, this could lead to multiple spikes in luminosity with approximately similar magnitudes that are clustered around times of periastron, but not exactly coincident with times of periastron. Multiple luminosity spikes may also be due to disconnected shells with cavities in between, perhaps analogous to the nested shells of dust around WR~140 that are created by repeated wind collisions \citep{L22}. In this case, luminosity spikes might happen at different times relative to periastron, depending on the exact location of a dense shell.  For example, if we presume that a brightness peak occurs when the companion encounters the densest CSM, and if circumstellar shells are thin, then the companion star may encounter the densest material somewhat before or after periastron.  Averaged over many orbits, the brightness peaks would still tend to be clustered around times of periastron, even if they do not occur exactly at periastron. 

As stated previously in Section~\ref{sec:obs}, the {\it Gaia} baseline magnitude of AT~2016blu has increased by $\sim 0.5$~mag over a period of $\sim 8$~yr. 
For the luminosity and temperature derived from the progenitor's photometry (see Section~\ref{sec:Precursor}), the corresponding radius is already near the maximum one would expect before strong interaction with a companion must start (i.e., the primary star's radius is comparable to the separation between the two stars).  If the slow brightening of the baseline quiescent flux from {\it Gaia} indicates a continued expansion of the LBV star, then this expansion may trigger increasingly more violent interaction.  Such expansion may also have initiated these eruptions a decade ago, with relatively weak or nonexistent interaction before then.

In the binary scenario explained above, the companion star may be either a main-sequence star or an evolved star.  If wind-wind interaction (i.e., colliding wind shocks) is the mechanism, both stars must be massive because the companion needs to have a strong wind.  Alternatively, the companion star may be a compact object, in which case accretion onto a neutron star or black hole may be an interesting potential power source for AT~2016blu's eruptions. X-ray observations are needed to confirm the evidence of accretion onto a compact companion.  Interestingly, in a blue straggler evolutionary scenario where AT~2016blu has evolved as a mass gainer (as discussed in the previous section), its companion --- the initially more massive primary that was the mass donor --- may have already exploded, perhaps leaving a compact object in an eccentric orbit.

The LBV-modulated binary interaction model proposed in our previous work has multiple free parameters such as orbital period, eccentricity, mass ratio, type of companion, etc.  Physical parameters can vary in each system and can lead to very different light curves that may nevertheless share the common attribute of having quasiperiodic, irregular bright eruptions.  For example, if AT~2016blu's orbit is somewhat less eccentric than that of SN~2000ch, then density fluctuations in the CSM (shells, disks) may cause the light curve to appear more irregular than that of SN~2000ch.

Using {\sl HST\/} images, we have determined that AT~2016blu should have a mass and luminosity higher than $M \approx 33$~M$_{\odot}$ and $L \approx 10^{5.7}$~L$_{\odot}$ (see Section~\ref{sec:Precursor} for details). The photospheric radius of such a star is $\sim 38$~R$_{\odot}$ (0.2~AU) when it is in a hot quiescent state with an effective temperature of $\sim 25,000$~K. However, if it is an LBV, then the photosphere should expand to $\sim 330$~R$_{\odot}$ (1.5~AU) during cooler outburst states with an effective temperature of $\sim 8500$~K. Based on the observed orbital period of $\sim 113$~d, the orbit's semimajor axis is estimated to be $a \approx 1.5 \, {\rm AU} \, (M_{\rm tot} / (M_{\rm primary}+M_{\rm secondary}))^{1/3}$. Here, $M_{\rm tot}$ represents the total mass of the system. However, as we do not currently have any information about the companion star, we assume it to be a star with a mass of 1~${\rm M}_{\odot}$. In other words, if the companion were a star with a mass of 10~${\rm M}_{\odot}$, the semimajor axis would be $a \approx 1.5 \, {\rm AU} \, (43 M_{\rm tot}/ 34M_{\rm tot})^{1/3}\approx$ 1.6 AU. If the orbit of AT~2016blu is less eccentric compared to SN~2000ch with an eccentricity of $\sim 0.4$, for instance, then the periastron separation would be about 0.9~AU and the apastron distance $\sim 2.1$~AU. 

At periastron, the distance between the two stars is smaller than the expanded radius of the LBV during its cooler eruptive state, which suggests a very strong interaction between the two stars. Conversely, at apastron, the distance between the stars is larger than the stellar radius, and interaction between the two stars may weaken or shut off. This could lead to cycles of interaction-driven outburst and quiescence. The intensity of the periastron interaction may also change slowly depending on the intrinsic state of the LBV. During some cycles, the interaction at periastron may be very weak, while at other times, we may expect more violent encounters as the LBV tries to expand.

As we mentioned above, the LBV in its outburst state is expected to have a radius of 1.5~AU, while the predicted periastron distance is only $\sim 0.9$~AU. In such a scenario, it may be possible for a companion star to briefly enter the LBV's stellar envelope, but it may not necessarily lead to the merging of the two stars owing to the small amount of mass contained within the outer envelope. However, a merger may eventually be triggered by the friction in these repeated encounters, so continued monitoring of AT~2016blu is worthwhile.  So far, we do not see strong evidence that the orbital period is changing.

There may be other mechanisms, beyond what we have proposed, that could account for the observed outbursts and periodicity of AT~2016blu without the need to invoke binarity. One alternative scenario might be that the behaviour is caused by an instability within a single massive star. 
Currently, however, we are aware of no predictions from single-star models that can account for quasiperiodic outbursts with a rapid and substantial change in luminosity comparable to those seen in AT~2016blu or SN~2000ch.

\section{Summary and Conclusion}\label{sec:conclusion}
We present the first analysis of photometric observations of AT~2016blu (= NGC~4559OT). AT~2016blu experienced its first known outburst in January 2012. Since that time, optical observations reveal that AT~2016blu experienced at least 19 known outbursts in 2012--2022.  Two of these outbursts in 2014 and 2016 are also detected in the IR by {\it Spitzer}. The outburst light curves show a wide variety of shapes, durations, and amplitudes. While the outbursts are irregular and multipeaked, our investigation demonstrates that they do repeat with a period of $\sim 113 \pm 2$~d.

We propose that AT~2016blu's outbursts are driven by interaction between two stars at periastron in an eccentric binary system, where the primary star is a massive LBV and the companion star may be a compact object or another star. Using {\sl HST\/} observations, we find that the mass and luminosity of AT~2016blu should be higher than $M \approx 33$~M$_{\odot}$ and $L \approx 10^{5.7}$~L$_{\odot}$, respectively.  These may be larger if there is circumstellar extinction that we have not taken into account. 

In a binary interaction scenario, the strength of periastron interaction between two stars may vary depending on the state of the LBV around the time of the periastron passage. An interaction with the companion at periastron may be relatively weak when the LBV is in a quiescent state with small stellar radius and low mass-loss rate. Stronger interactions are expected, however, in periastron passes when the LBV has an inflated radius during an S~Dor excursion and perhaps a higher mass-loss rate. Thus, successive periastron eruptions may look different each time.

The proposed binary interaction of AT~2016blu is similar to the periastron encounters envisioned for SN~2000ch \citep{AA22},
SN~2009ip \citep{P13,S14,S22}, and $\eta$~Car \citep{SN11,S18}. The outbursts of AT~2016blu are less regular and less periodic than those of SN~2000ch, indicating that the AT~2016blu system may have a less eccentric orbit with significantly clumpy CSM, or multiple expanding dust shells, or unsteady accretion onto the compact companion. Further studies are required to understand the nature of the companion and underlying physics causing multiple spikes in the light curve.

SN~2000ch and AT~2016blu are so far the only studied extragalactic LBVs that exhibit quasiperiodic behaviour in their major outburst variability. According to some studies in the literature, AG~Car in the Milky Way, as well as LMC LBV S~Dor, exhibit quasiperiodic behaviour, and $\eta$~Car is an eccentric colliding-wind binary with a period of 5.5~yr  \citep{v97, D08}. However, these estimated periods are longer than those of AT~2016blu and SN~2000ch, and are associated with the small-amplitude variability of LBVs, or (in the case of $\eta$~Car) colliding winds that have little impact on the total luminosity.

Given the detected period of $\sim 113$~d, we predict that AT~2016blu's next outburst will be around late February/March 2023\footnote{This date is before the publication date, but well after the original submission date.  At the time of resubmitting this paper, we do not analyse data covering this event.} and then again in late-June 2023, October 2023, and in February 2024, although the event in October 2023 is likely to be unobservable. Since AT~2016blu's light curve shows that the outbursts are usually brief, multipeaked, and sometimes occur before or after the expected time of periastron, it would be worthwhile to monitor AT~2016blu with high-cadence photometry and spectroscopy for about a month before and after the proposed timelines. We encourage further monitoring and detailed observations of AT~2016blu considering that it may be a prelude to a merger or SN explosion, similar to SN~2009ip.

\section*{Acknowledgements}
We thank the anonymous referee for helpful comments that improved the quality of this paper. This research has made use of the NASA/IPAC Infrared Science Archive, which is funded by the National Aeronautics and Space Administration (NASA) and operated by the California Institute of Technology. It also used data from the Asteroid Terrestrial-impact Last Alert System (ATLAS) project, which is funded primarily to search for near-Earth objects (NEOs) through NASA grants NN12AR55G, 80NSSC18K0284, and 80NSSC18K1575; byproducts of the NEO search include images and catalogues from the survey area. The ATLAS science products have been made possible through the contributions of the University of Hawaii Institute for Astronomy, the Queen’s University Belfast, the Space Telescope Science Institute (STScI), the South African Astronomical Observatory, and The Millennium Institute of Astrophysics (MAS), Chile. We also acknowledge ESA {\it Gaia}, DPAC, and the Photometric Science Alerts Team (http://gsaweb.ast.cam.ac.uk/alerts). This work is also based on observations made with the {\it Spitzer Space Telescope}, which is operated by the Jet Propulsion Laboratory, California Institute of Technology, under a contract with NASA.

This work was partially funded by Kepler/K2 grant J1944/80NSSC19K0112, as well as by STFC grants ST/T000198/1 and ST/S006109/1. Financial support was provided by NASA through {\it HST} grants AR-14259, AR-14316, and GO-15889 from STScI, which is operated by the Association of Universities for Research in Astronomy (AURA), Inc., under NASA contract NAS5-26555. A.V.F.'s group at UC Berkeley has received generous financial assistance from the Christopher R. Redlich Fund, Alan Eustace (W.Z. is a Eustace Specialist in Astronomy), and numerous individual donors. KAIT and its ongoing operation were made possible by donations from Sun Microsystems, Inc., the Hewlett-Packard Company, AutoScope Corporation, Lick Observatory, the U.S. National Science Foundation (NSF), the University of California, the Sylvia \& Jim Katzman Foundation, and the TABASGO Foundation.  Research at Lick Observatory is partially supported by a generous gift from Google. J.E.A.\ is supported by the international Gemini Observatory, a program of NSF's NOIRLab, which is managed by AURA, Inc., under a cooperative agreement with the NSF, on behalf of the Gemini partnership of Argentina, Brazil, Canada, Chile, the Republic of Korea, and the USA.

\section*{Data Availability}
The KAIT, Super-LOTIS, Kuiper, and {\it Spitzer} data used in this work are available in the article.  ZTF data are available in the public domain \url{https://irsa.ipac.caltech.edu/Missions/ztf.html}. ATLAS data are available in the ATLAS Forced Photometry server \url{https://fallingstar-data.com/forcedphot/}. {\it Gaia} data are available in \url{http://gsaweb.ast.cam.ac.uk/alerts/home}. The ``Bright Supernovae" data are available at \url{https://www.rochesterastronomy.org/supernova.html}.

\bibliographystyle{mnras}
\DeclareRobustCommand{\VAN}[3]{#3}
\bibliography{MA}
\label{lastpage}

\appendix
\section{The Photometric Data}
The photometric data used in this work are presented in the following tables. All photometric data  presented below are in the Vega system.

\begin{table}
\caption{Unfiltered KAIT Photometry ($\sim R$ band).}\scriptsize
\centering
\label{tab:KAIT}
\begin{tabular}{lccc} 
\hline
JD& Mag& $\sigma$  \\
\hline
2455936.04&19.24&0.28\\
2455938.04&16.78&0.04\\
2455938.91&16.69&0.04\\
2455940.05&17.73&0.07\\
2455942.03&18.55&0.28\\
2456048.77&18.30&0.12\\
2456076.74&17.45&0.08\\
2456327.88&17.70&0.08\\
2456713.93&18.04&0.09\\
2457387.08&17.62&0.10\\
2459371.74&17.64&0.06\\
2459357.81&17.86&0.11\\
2459358.81&17.83&0.12\\
2459324.90&17.10&0.04\\
2459326.89&17.17&0.06\\
\hline
\end{tabular}
\end{table}

\begin{table}
\caption{Super-LOTIS Photometry.}\scriptsize
\begin{threeparttable}
\centering
\begin{adjustbox}{max width=1.1\textwidth,center}
\begin{tabular}{lccccccccccc} 
\hline
JD& $R$ & $\sigma_{R}$ & $V$ &$\sigma_{V}$ & $I$ &$\sigma_{I}$\\
\hline
2457879.80&18.35&0.06&18.35&0.05&  &  \\
2457885.80&18.45&0.07&18.32&0.07&  &  \\
2457893.80&17.82&0.02&17.94&0.02&  &  \\
2457895.80&18.09&0.05&18.19&0.05&  &   \\
2458204.90&18.59&0.07&  &  &16.61&0.41\\
2458213.90&17.48&0.03&  &  &17.68&0.04\\
2458229.90&18.39&0.04&  &  &18.87&0.06\\
2458242.80&18.51&0.04&  &  &  &  \\
2458246.80&18.42&0.04&  &  &  &  \\
2458255.80&18.52&0.04&  &  &  &  \\
2458257.80&18.67&0.04&  &  &  &  \\
2458258.80&18.42&0.04&  &  &  &  \\
2458263.60&18.56&0.04&  &  &  &  \\
2458266.60&18.93&0.09&  &  &  &  \\
2458269.60&18.66&0.06&  &  &  &  \\
2458457.00&17.49&0.06&17.41&0.05&  &  \\
2458473.00&18.48&0.10&18.76&0.09&  &  \\
2458567.00&18.47&0.07&  &  &  &  \\
2458606.80&18.57&0.05&18.57&0.04&  &\\  
2458618.80&18.89&0.11&  &  &  &  \\
2458638.80&17.22&0.03&  &  &  &  \\
2458640.80&18.14&0.04&  &  &  &  \\
2458645.70&18.15&0.09&  &  &  &  \\
2458651.80&17.27&0.34&  &  &  &  \\
2458654.70&18.40&0.05&  &  &  &  \\
2458660.70&18.61&0.04&  &  &  &  \\
2458871.90&18.41&0.02&  &  &  &  \\
2458886.90&18.35&0.03&  &  &  &  \\
2458894.90&18.47&0.02&  &  &  &  \\
2459525.00&18.38&0.04&  &  &  &  \\
2459531.00&18.12&0.03&  &  &  &  \\
2459554.00&18.40&0.02&  &  &  &  \\
2459560.00&17.21&0.01&  &  &  &  \\
2459566.00&17.81&0.06&  &  &  &  \\
2459586.00&18.32&0.02&  &  &  &  \\
2459592.00&18.19&0.03&  &  &  &  \\
2459601.00&17.49&0.02&  &  &  &  \\
2459710.80&18.08&0.04&18.01&0.04&  &\\  
2459733.70&18.34&0.03&18.59&0.03&  &  \\

\hline
\end{tabular}
\end{adjustbox}
\end{threeparttable}
\label{tab:Super-LOTIS}
\end{table}

\begin{table}
\caption{Kuiper/MONT4K Photometry.}\scriptsize
\centering
\label{tab:Kuiper}
\begin{tabular}{lccc} 
\hline
JD& Harris-$R$ & $\sigma_{\rm{R}}$ \\
\hline
2457466.77&18.82&0.05\\
2457520.81&19.28&0.06\\
2457535.79&19.13&0.08\\
2457805.88&18.84&0.08\\
2457900.80&18.39&0.12\\
2457923.77&19.03&0.07\\
2458108.92&18.54&0.07\\
2458227.79&18.73&0.15\\
2458257.83&19.35&0.07\\
2458582.78&18.87&0.24\\
2458612.80&19.19&0.14\\
2459257.82&18.69&0.09\\
2459321.85&18.49&0.02\\
2459322.83&17.20&0.13\\
2459345.88&18.45&0.11\\
2459639.98&17.09&0.02\\
2459640.90&17.68&0.10\\
2459641.80&17.80&0.14\\
2459666.76&18.91&0.05\\
2459706.79&18.46&0.03\\
2459707.82&18.31&0.12\\
2459903.01&16.41&0.14\\
2459984.78&18.26&0.12\\
2459985.82&18.06&0.09\\

\hline
\end{tabular}
\end{table}

\begin{table}
\caption{{\sl HST\/} Raw Photometry.}\scriptsize
\centering
\label{tab:HST}
\begin{tabular}{lccccc} 
\hline
JD & HST mode& Spectral Element & Mag &  $\sigma$\\
\hline
2452054.5& WFPC2& F450W &20.552& 0.006\\
&&F555W& 20.427& 0.005\\
&&F814W& 19.920& 0.005\\

2453437.5& ACS/HRC& F435W& 20.489& 0.003\\
&&F555W& 20.336 & 0.003\\
&&F814W&19.932 & 0.003\\

\hline
\end{tabular}
\end{table}

\begin{table}
\caption{Spitzer/IRAC Photometry.}\scriptsize
\centering
\label{tab:spitzer}
\begin{tabular}{lcccccc} 
\hline
JD& Mag$_{[3.6]}$& $\sigma_{[3.6]}$ & Mag$_{[4.5]}$& $\sigma_{[4.5]}$& Colour$^*$ \\
\hline
2453148.87&17.05&0.06&16.90&0.08&...\\
2453152.21&17.13&0.07&16.91&0.09&...\\
2456723.28&16.66&0.04&16.34&0.04&0.49\\
2456751.17&16.98&0.05&16.74&0.06&0.43\\
2456906.66&17.04&0.05&16.83&0.07&0.14\\
2457092.01&17.09&0.05&16.84&0.06&1.06\\
2457251.25&16.92&0.05&16.56&0.04&0.92\\
2457466.48&16.88&0.05&16.60&0.05&0.51\\
2457472.39&16.63&0.04&16.28&0.03&0.55\\
2457487.16&16.86&0.04&16.45&0.04&0.89\\
2457619.52&17.08&0.06&16.76&0.05&1.63\\
2457627.15&17.13&0.06&16.91&0.07&...\\
2457648.27&17.10&0.05&16.89&0.06&0.17\\
\hline

\end{tabular}
\begin{tablenotes}
   \item[*] $^*$After subtracting the minimum observed flux before 2012.
  \end{tablenotes}
\end{table}

\end{document}